\documentclass[final,5p,times,twocolumn,authoryear]{elsarticle}
\usepackage{amssymb,amsmath}
\usepackage{amsthm}
\usepackage{jtb}
\usepackage{color}

\newcommand{\cC}{{\cal C}}
\newcommand{\tc}{{\tilde c}}

\newcommand{\permu}[2]{ ( #1  )_#2}
\newcommand{\sref}[1]{Sec.~\ref{#1}}
\newcommand{\A}[1]{A\left ( \{\sigma\}_{#1} \right )}
\journal{Journal of Theoretical Biology}

\begin{document}

\begin{frontmatter}

\title{Greedy adaptive walks on a correlated fitness landscape}
\author[kor]{Su-Chan Park\corref{cor1}}
\ead{spark0@catholic.ac.kr}
\author[ger]{Johannes Neidhart}
\author[ger]{Joachim Krug}
\cortext[cor1]{Tel:+82-2-2164-4524, Fax: +82-2-2164-4764}

\address[kor]{Department of Physics, The Catholic University of Korea, Bucheon 14662, Republic of Korea}
\address[ger]{Institut f\"ur Theoretische Physik, Universit\"at zu K\"oln, 50937 K\"oln, Germany}

\begin{abstract}
We study adaptation of a haploid asexual population on a fitness landscape defined over binary genotype sequences of length $L$. We consider greedy adaptive walks in which the population moves to the fittest among all single mutant neighbors of the current genotype until a local fitness maximum is reached. The landscape is of the rough mount Fuji type, which means that the fitness value assigned to a sequence is the sum of a random and a deterministic component. The random components are independent and identically distributed random variables, and the deterministic component varies linearly with the distance to a reference sequence. The deterministic fitness gradient $c$ is a parameter that interpolates between the limits of an uncorrelated random landscape ($c = 0$) and an effectively additive landscape ($c \to \infty$). When the random fitness component is chosen from the Gumbel distribution, explicit expressions for the distribution of the number of steps taken by the greedy walk are obtained, and it is shown that the walk length varies non-monotonically with the strength of the fitness gradient when the starting point is sufficiently close to the reference sequence. Asymptotic results for general distributions of the random fitness component are obtained using extreme value theory, and it is found that the walk length attains a non-trivial limit for $L \to \infty$, different from its values for $c=0$ and $c = \infty$, if $c$ is scaled with $L$ in an appropriate combination. 

\end{abstract}

\begin{keyword}
Adaptation \sep Genotype space
\sep Extreme value theory \sep Asexual population

\end{keyword}

\end{frontmatter}

% \linenumbers

\section{\label{Sec:intro}Introduction}
Ever since the concept of the \textit{fitness landscape} was introduced by
Sewall \citet{W1932}, it has played a central role in evolutionary
biology~\citep{deVisser2014}. 
Among the different variants of the concept used in the literature, 
we here restrict ourselves to fitness landscapes that 
%The fitness landscape is 
map the genotype space into the
real numbers by assigning a fitness value to every genotype.
With this definition, the fitness landscape provides an intuitive
picture of evolution as a hill-climbing process. A convenient choice
for the genotype space is the $L$-dimensional hypercube $\{0,1\}^L$, which contains all binary sequences $\cC =
(1,0,1,\dots,1,1)$ of length $L$. Rather than specifying the genome
on the level of DNA base pairs, the binary sequences keep track of the
presence or absence of mutations compared to a wild-type genome, or
(in a more coarse-grained representation) the presence or absence of entire genes.

In addition to the underlying fitness landscape,
the dynamics of adaptation is governed by the population size $N$ and the
mutation rate $U$ per genome, both of which are to be compared to the 
scale of fitness differences summarized in a typical selection coefficient $s$. In the
\textit{strong selection / weak mutation} (SSWM) regime characterized
by the conditions $Ns\gg 1$ and $NU\ll1$ the population is monomorphic for
most of the time, and the adaptive process is guided by the landscape
structure in a simple way~\citep{G1983,G1984}. If
a mutation to a fitter genotype occurs it has a nonzero probability of
fixing in the population, whereas a mutation to a sequence with lower fitness
is certain to go extinct. The low mutation rate makes it very unlikely for double
mutations to occur. Accordingly, in this regime the population behaves
as a point in sequence space that moves uphill in the fitness
landscape by single mutational steps, 
a process referred to as an \textit{adaptive walk}~\citep{G1983,G1984,KL1987}.
An obvious feature of adaptive walks is that they end on a
 fitness maximum, that is, a genotype without fitter one-mutant neighbors.
 This makes the \textit{walk length}, the number of steps until a maximum is
 reached, a property of interest. 
 
A simplified version of the adaptive walk problem where 
the effect of mutant fitness on the fixation probability of
beneficial mutations is neglected and any neighboring genotype of higher fitness
can fix with equal probability was studied by~\cite{Macken1989}
and~\cite{FL1992}. For rugged landscapes without fitness correlations 
the mean number of steps of such `random' adaptive walks was found to
be of the order of $\ln L$.
When the effect of the fixation probability is incorporated the mean
walk length is still logarithmic in the number of loci, but the coefficient of $\ln
L$ becomes dependent on the distribution of fitness values~\citep{G1983,O2002,NK2011,Jain2011,Seetharaman2011}.
If the infinite $L$ limit is taken the walks no longer terminate and
adaptation can be studied through the unbounded
increase of the mean fitness of the population~\citep{PK2008}.

When the population size is increased beyond the SSWM regime, the number of segregating
sites becomes larger than two. In asexual populations this implies that two beneficial mutations compete with each
other for fixation and the one with the larger fitness will be fixed preferentially.
This phenomenon is connected to the Hill-Robertson
effect~\citep{HR1966} and is commonly known as
\textit{clonal interference}~\citep{GL1998,W2004,PK2007,DF2007,PSK2010}.
A rough criterion for the clonal interference regime is provided by the
condition $N U \ln N \gg 1$.
If we denote the mean fixation time in this regime by $T_0$ (which depends on $N$, $U$, and $s$),
almost all beneficial single mutant neighbors of the most populated
genotype will be present during the fixation process if $N U T_0 \gg L$.
To model this regime by an adaptive walk, we use a deterministic rule for the
next step: the walker chooses the 
genotype with the largest fitness among the sequences that are one mutation away.
This kind of adaptive walk was called a `perfect' or `gradient'
adaptive walk by~\cite{O2002,O2003}, but here we follow \citet{KL1987}
in referring to it as a \textit{greedy} walk.
\citet{O2003} calculated the length of a greedy adaptive
walk on an uncorrelated fitness landscape using an order statistics approach
that is independent of the fitness distribution, provided it is continuous.
In the limit $L\to \infty$ the mean walk length is given by $e -1 \approx 1.72$,
which was suggested to be a lower bound on the mean number of
steps for any adaptive walk [see also \citet{R2005}]. Note that for
this description to faithfully represent adaptation under
strong clonal interference, the mutation rate has to be small enough
such that the creation of double mutants can be neglected
\citep{SFVK2013}. 

The studies of adaptive walks mentioned above were based on the
assumption of an uncorrelated random fitness landscape with maximal 
ruggedness, which is not supported by empirical
evidence~\citep{Miller2011,SSFKV2013,deVisser2014}. The effect of
fitness correlations on adaptive walks has so far been addressed mostly
in the context of `block model' landscapes in which the genotype
is subdivided into independent modules, each of which is assigned a
random fitness, and the mean walk length is additive over 
modules~\citep{Perelson1995,O2006,Seetharaman2014, Nowak2015}. Here we consider
greedy adaptive walks on another class of tunably rugged fitness
landscapes, the rough mount Fuji (RMF) model, which was originally 
introduced in the context of protein evolution~\citep{AUINKH2000}.
In the RMF model an uncorrelated random fitness landscape is
superimposed on a linear fitness gradient, and the slope of this
gradient serves as a tuning parameter controlling the ruggedness of
the landscape.

The RMF model has recently been found to provide a convenient parametrization
of many empirical fitness data sets~\citep{FKdVK2011,SSFKV2013,NSK2013}, while at the same time
allowing for detailed mathematical analysis of a wide range of
landscape properties~\citep{NSK2014,PSNK2014}.
Of particular interest for our work are the results on the existence
of selectively accessible mutational pathways, defined here as
pathways to the global fitness maximum along which fitness increases
monotonically and which are moreover \textit{directed}, in the sense
that the distance to the global optimum decreases in each
step~\citep{WWC2005,FKdVK2011}. 
\citet{HM2014} have shown that such pathways exist in the RMF model with a probability
approaching unity for $L \to \infty$, whereas this probability
tends to zero for uncorrelated landscapes. A population following a directed accessible
pathway would perform an adaptive walk of ${\cal{O}}(L)$ steps,
much longer than the walks on uncorrelated landscapes.  
However, the biological significance of accessible paths is not evident,
because an evolving population may not find them even if they exist~\citep{SFVK2013,PSNK2014}.

In this paper, we study greedy adaptive walks on the RMF fitness landscape, focusing
on the mean number of steps when $L$ is very large. For a specific
choice of the distribution of the random fitness component in the RMF
model we obtain an analytic solution for the full distribution of walk
lengths and show that it attains a non-degenerate limit for $L \to
\infty$, similar to Orr's analysis of the uncorrelated case~\citep{O2003}. 
We also consider the dependence of the walk length
on the distance of the initial genotype from the \textit{reference state}, 
and show that in a range of distances the walk length
varies non-monotonically with the strength of the fitness gradient. 

Arbitrary distributions of the random fitness
component can be treated in the limit $L \to \infty$ by exploiting the
convergence of the maximum of $L$ random variables to one of the
universal distributions of extreme value theory (EVT)~\citep{deHaan}.
The EVT approach to adaptation 
was pioneered by \cite{G1984} and \cite{O2002} and has meanwhile become an established conceptual framework that allows to 
organize and quantify the relation between the distribution of mutational effects and the corresponding adaptive 
behavior \citep{Joyce2008,O2010,Rokyta2008,Schenk2012,Bank2014}. 
Similar to the analysis of
fitness landscape properties for the RMF model presented by \cite{NSK2014},
we find that the behavior of the walk length is governed by the
interplay between the ruggedness parameter and the tail properties of
the distribution of the random fitness component. Specifically, if the
tail of the distribution is fatter than exponential, the walk length
reverts to the behavior found by Orr for uncorrelated landscapes for
any fixed value of the fitness gradient. On the other hand, for tails thinner than exponential the effective
strength of the fitness gradient increases without bound with increasing $L$, such that the greedy
walks traverse the entire landscape with high probability for $L \to \infty$. 
A non-trivial limit of the walk length is attained only when $c$ and $L$ 
are scaled together in a particular combination. 

\section{\label{Sec:RMF} Definitions}
The RMF fitness landscape is constructed from an additive `mount Fuji' fitness landscape by adding an independent
and identically distributed (i.i.d.) random variable to the fitness of
every genotype. By $\cC$ we denote
a binary sequence of length $L$ which represents the genotype. In particular,
we will call the sequence $\cC_r = \left ( 1, 1, \ldots, 1 \right )$ the
\textit{reference sequence}  which has the
largest fitness in the purely additive landscape. Its antipodal point
on the hypercube, the sequence with all elements 0, will be denoted by 
$\cC_a$.
The fitness of a sequence $\cC$ in the RMF fitness landscape is then assigned as
\begin{equation}
\label{Eq:RMF}
W(\cC) = - c d_r(\cC) + \xi_\cC,
\end{equation}
where $d_r(\cC)$ is the Hamming distance between $\cC$ and the
reference sequence $\cC_r$, $c$ is a positive real number, and $\{\xi_\cC\}$ are i.i.d. random variables with
probability density $f(\xi)$ and cumulative distribution function $F(\xi)$, defined
as
\begin{equation}
F(\xi) = \int_{-\infty}^\xi f(x) \, dx.
\end{equation}
The definition (\ref{Eq:RMF}) should be interpreted in the Malthusian sense, where fitness values can be positive 
or negative. What \cite{HM2014} proved is that for $c>0$ in the limit $L\to \infty$ there is almost surely a 
directed path from the antipode $\cC_a$ to the reference
sequence $\cC_r$ along which fitness is monotonically increasing, irrespective
of the actual form of $f(\xi)$, whereas for $c=0$ such paths almost
surely do not exist.

Since we are interested in greedy walks, the statistics of the maximal
value among groups of i.i.d. random variables will play an important
role. For this reason we introduce the probability $G_k(x)$ that 
the largest value among $L-k+1$ ($k\ge 1$) i.i.d. $\xi$'s is smaller than $x$,
which is
\begin{equation} 
G_k(x) \equiv \left ( \int_{-\infty}^x f(y) dy \right )^{L-k+1}
= F(x)^{L-k+1},
\label{Eq:Gk}
\end{equation}
with the corresponding density $g_k(x)$ 
\begin{equation}
g_k(x) = (L-k+1) f(x) F(x)^{L-k}.
\label{Eq:gk}
\end{equation}
The reason for considering $L-k+1$ variables rather than $k$ variables will become
clear in \sref{Sec:mean}.

As has been noted previously~\citep{Franke2010,FKdVK2011,NSK2014},
many properties of the RMF model take on a particularly simple form
when the random variables $\xi_{\cC}$ are drawn from the Gumbel
distribution $f(x) = e^{-x - e^{-x}}$, and we will adopt this choice
in \sref{Sec:mean}. For the Gumbel distribution, $G_k(x)$ and $g_k(x)$ become
\begin{align}
\label{Eq:gkGum}
g_k(x) &= (L-k+1) e^{-x - (L-k+1) e^{-x}}, \\ 
G_k(x) &= \int_{-\infty}^x g_k(x) = e^{-(L-k+1) e^{-x}}.
\label{Eq:GkGum}
\end{align}
The Gumbel distribution is one of the three universal limiting
distributions that arise in extreme value theory \citep{deHaan}, and we will exploit
this connection in \sref{Sec:bound} where we study the properties of
greedy adaptive walks for general choices of the distribution $f(x)$.

\section{\label{Sec:mean}Gumbel-distributed random
  fitness component}

\subsection{\label{Sec:antipodal}Greedy walks starting from the
  antipodal sequence}

Our analysis begins with the greedy walk starting from the antipodal sequence $\cC_a$.
As mentioned before, the probability that at least
one accessible path from $\cC_a$ to $\cC_r$ exists converges to unity as
$L\to\infty$ for any finite $c>0$~\citep{HM2014}. If the greedy walker takes such a path with probability of $O(1)$,
the mean number of steps will be $O(L)$.
On the other hand, the RMF with $c=0$ is identical to the uncorrelated rugged landscape
or the House-of-Cards model~\citep{K1978} and the mean number of steps of greedy walks is 
$e-1\simeq 1.72$ in the limit of infinite $L$~\citep{O2003}. Thus the
first question to address is whether the greedy walk length remains finite
for $L \to \infty$ when $c > 0$.

\subsubsection{Exact solution}
To find the mean walk distance, we consider 
the probability $H_l$ that the walker takes at least $l$ steps. 
For convenience, we denote the sequence at the $l$-th step by $\cC_l$
with $\cC_0 = \cC_a$.
The fitness of $\cC_l$ is the largest among the single mutant neighbors of $\cC_{l-1}$.
To find $H_l$, we make the assumption that $d_r(\cC_l)$ is a decreasing function in $l$, that is, 
the walker only takes steps in the direction towards the reference
sequence $\cC_r$, referred to as the \textit{uphill} direction in the following. 
This assumption is plausible if $L\gg l$, because a \textit{downhill} step is possible
only if the largest among the $l$ random fitness components  of the downhill neighbors exceeds the largest among the $L-l$ random fitness components of the uphill neighbors by at least $2c$ . Obviously, for
reasonably large $L$ and a setting with rather short walks, this
probability is negligible.
The validity of this assumption will be ascertained
later in a self-consistent way. Once the $H_l$ have been determined, 
it follows that the greedy walk takes exactly $l$ steps with
probability $H_l-H_{l+1}$
and, in turn, the mean number of steps is
\begin{equation}
\label{Eq:deflength}
\left \langle l \right \rangle 
= \sum_{l=1}^{L} l \left (  H_l - H_{l+1}  
\right )  
=  \sum_{l=1}^L H_l,
\end{equation}
where $H_{L+1}$ is set to $0$.

Let $J_l(x)$ be the probability that the walker takes at least $l$ steps with $W(\cC_l) < -c(L-l) + x$ and let $j_l(x) = \frac{d}{dx} J_l(x)$
($l=0,1,\ldots, L$).
Obviously, 
\begin{align}
\label{Hldef}
H_l = \lim_{x\rightarrow\infty} J_l(x) = \int_{-\infty}^\infty j_l(y) dy.
\end{align}
A recursion relation for $j_l(x)$ can be derived immediately from the
definition:
\begin{equation}
\label{basicrec}
j_l(x) = g_l(x) J_{l-1}(x+c)
= g_l(x) \int_{-\infty}^{x+c} dy j_{l-1}(y)
\end{equation}
with $j_0(x) = f(x) = e^{-x - e^{-x}}$.
Since $\cC_{l-1}$ has $L-l+1$ nearest neighbors in the uphill direction, we have considered
$g_l(x)$ defined in Eq.~\eqref{Eq:gk} in the recursion relation.

Introducing 
\begin{align}
a_k &= e^{-kc} + \sum_{m=0}^{k-1}(L-k+m+1) e^{-mc}\nonumber\\
&= \frac{L \left(1-e^{- k c}\right)-e^{- (k+1)c}-k}{1-e^{-c}}
+\frac{1-e^{- (k+1) c}}{\left(1-e^{-c}\right)^2}
\end{align}
which satisfies the recursion relation $a_{k+1} = (L-k) + e^{-c} a_k$ with $a_0=1$,
we can write
\begin{equation}
j_l(x) = \frac{L!}{(L-l)!} \left ( \prod_{k=0}^{l-1} \frac{1}{a_k} \right )
e^{-x - a_l e^{-x}},
\label{Eq:jl}
\end{equation}
for $l\ge 1$, which can be proved straightforwardly by mathematical induction.
Thus, we get 
\begin{equation}
H_l = \prod_{k=1}^{l} \frac{L-k+1}{a_k} 
\label{Eq:Hl}
\end{equation}
as an exact expression for the distribution of walk length.
Note that in the above derivation, the sign of $c$ does not play any role,
which implies that the case of negative $c$ can be studied within the
same scheme and Eq.~\eqref{Eq:Hl} is valid for any $c$. 
By symmetry, a greedy walk with negative $c$ can be interpreted as a walk starting from
the reference sequence $\cC_r$ (see \sref{Sec:arbid} for further discussion).

Since it does not appear feasible to extract simple analytic formulae
from (\ref{Eq:Hl}) for arbitrary $c$ and $L$, below   
we will present approximate calculations for certain limiting cases.
Before delving into detail, we derive a simple upper bound on $\langle l \rangle$.
Since $a_k \ge (L-k+1) + (L-k+2) e^{-c} \ge (L-k+1) ( 1 + e^{-c} )$ for $k\ge 2$ and $a_1 = L+e^{-c} \ge L$,
\begin{equation}
H_l 
\le (1+e^{-c})^{-(l-1)},
\label{Eq:Hlupper}
\end{equation}
which gives 
\begin{equation}
\langle l \rangle = \sum_{l=1}^L H_l \le \sum_{l=1}^\infty (1+e^{-c})^{-(l-1)} = 1+e^c.
\label{Eq:upper}
\end{equation}
This upper bound clearly shows that $\langle l \rangle / L \rightarrow
0$ as $L \rightarrow \infty$ for any $c$ when $\xi_C$ is drawn from the Gumbel distribution.
That is, it is highly unlikely that a greedy walk can follow 
an accessible path all the way to the reference state, although such paths exist with probability 1
as shown by \cite{HM2014}.

\subsubsection{The limit $L \to \infty$ at finite $c$}

Since Eq.~\eqref{Eq:Hlupper} is valid for any $L$, $H_l$ should be exponentially small for
$l \sim O(L)$ once $L \gg e^c$. 
This self-consistently affirms the validity of the assumption used in
writing down $H_l$. In order to extract the $L \to \infty$ limit of
$H_l$ from (\ref{Eq:Hl}) , we use
$a_k \sim L (1-e^{-kc})/(1-e^{-c})$ to obtain 
\begin{equation}
H_l = \prod_{k=1}^l \frac{1-e^{-c}}{1 - e^{-kc}}.
\label{Eq:HlLinf}
\end{equation}
This expression has an appealing interpretation in terms of so-called
$q$-analogues \citep{Koekoek}.
Recall that the $q$-analogue of a number $n$ can be defined by $[n]_q
= (1-q^n)/(1-q)$, which satisfies the basic property that $\lim_{q
  \to 1} [n]_q = n$. Defining the $q$-factorial as $[n]_q! =
\prod_{k=1}^n [k]_q$, we see that 
$H_l = ([l]_{e^{-c}}!)^{-1}$
which reduces to Orr's result $H_l = (l!)^{-1}$ in the limit $c \to
0$, $e^{-c} \to 1$. Moreover, the mean walk length is given by 
\begin{equation}
\label{Eq:lq}
\langle l \rangle = \sum_{l=1}^\infty H_l = \exp_{e^{-c}}(1)-1,
\end{equation}
where $\exp_q(x)$ is the $q$-exponential function, defined as
$$
\exp_q(x) = \sum_{n=0}^\infty \frac{1}{[n]_q!} x^n. 
$$
In fact an
alternative derivation of (\ref{Eq:HlLinf}) can be set up in complete
analogy to the original approach of \cite{O2003} [see \cite{Neidhart2014}]. 

We note for later reference that the expression (\ref{Eq:HlLinf})
has been derived previously for the probability that $l$ 
random variables $y_k = x_k + ck$ are ascendingly ordered, $y_1 < y_2
< ...< y_l$, where the $x_k$ are drawn independently from a Gumbel
distribution \citep{Franke2010}. The reason for this coincidence will become
clear below in \sref{Sec:genintro}.

\subsubsection{Approximations for large and small $c$}
\label{Sec:largesmallc}

We next evaluate (\ref{Eq:HlLinf}) for large and small $c$, respectively.
If $c \gg 1$, $H_l$ can be approximated as
\begin{equation}
H_l \approx \frac{(1-e^{-c})^{l-1}}{1-e^{-2c}}
\end{equation}
for $l \ge 2$ and $H_1 = 1$.
In the above approximation, we have kept terms up to $O(e^{-2c})$ 
in the denominator.
Hence the mean distance becomes
\begin{equation}
\langle l \rangle \approx 1 + \sum_{k=2}^\infty \frac{(1-e^{-c})^{k-1}}{1-e^{-2c}}
= \frac{e^{c} - e^{-2c}}{1 - e^{-2c}} \approx e^c + e^{-c} 
\label{Eq:cgg1}
\end{equation}
which is close to the upper bound of Eq.~\eqref{Eq:upper}.

For $|c| \ll 1$, we expand $(1-e^{-c})/(1-e^{-kc})$  up to $O(c^3)$, which yields
\begin{align}
\frac{1-e^{-c}}{1-e^{-kc}} 
\approx \frac{1}{k}
\exp\left ( \frac{k-1}{2} c  - \frac{k^2-1}{24} c^2 + O(c^4)\right ).
\end{align}
Accordingly, $H_l$ is approximated as
\begin{align}
l! H_l \approx& \exp\left ( \frac{l(l-1)}{4} c - (2 l^3 + 3 l^2 - 5 l) \frac{c^2}{144} + O(c^4)\right ) \nonumber \\
=& 
1 + \frac{\permu{l}{2}}{4} c + \frac{9 \permu{l}{4} + 32 \permu{l}{3}}{288} c^2 \nonumber \\
&+ \frac{3\permu{l}6 + 32 \permu{l}5 + 72 \permu{l}4 - 24 \permu{l}2}{1152} c^3 +O(c^4),
\end{align}
where the Pochhammer symbol $\permu{l}k = l!/(l-k)!$ has been used.
Since $\sum_{l=1}^\infty \permu{l}k / l! = e -\delta_{k,0}$,
the mean distance becomes
\begin{equation}
\langle l \rangle = e \left ( 1 + \frac{1}{4}c + \frac{41}{288}c^2 + \frac{83}{1152} c^3 \right ) -1,
\label{Eq:smallc}
\end{equation}
which reproduces the result by \cite{O2003} when $c=0$. The fact that
the leading order correction is linear in $c$ implies that walks
starting at the references sequence ($c < 0$) are \textit{shorter}
than $e-1$ when $\vert c \vert$ is small. We will see below in \sref{Sec:arbid}
how this result generalizes to walks starting close to, but not at the
reference sequence.

If $c<0$ and $|c| \gg 1$, $(L-k+1)/a_k\approx e^{-(k-1)|c|}$ and 
$H_l = O(e^{-l(l-1)|c|/2})$.
Hence, to keep terms up to order $O(e^{-2|c|})$, it is enough to consider
only $H_1+H_2$, which gives
\begin{equation}
\langle l \rangle = 1+ e^{-|c|} - e^{-2|c|} + O(e^{-3|c|}).
\label{Eq:mcgg1}
\end{equation}
Note that even if $|c| \rightarrow \infty$, the walker takes at least one step.
This is because we take $L\rightarrow \infty$ limit before $|c| \rightarrow
\infty$ limit and under this order of limits the probability that the reference sequence is a local maximum is zero for any 
$c$. For later purposes we recall that the probability for a sequence
at distance $d$ from the reference sequence to be a local fitness
maximum is given by  \citep{NSK2014}
\begin{equation}
\label{Eq:maxprob} 
p_c^\mathrm{max}(d) = \frac{1}{1 + d e^c + (L-d) e^{-c}}
\end{equation}
which vanishes when the limit $L \to \infty$ is taken for $d=0$ and fixed $c$.
Thus the walker needs to take at least one step to
reach a maximum.

\begin{figure}[t]
\includegraphics[width=\columnwidth]{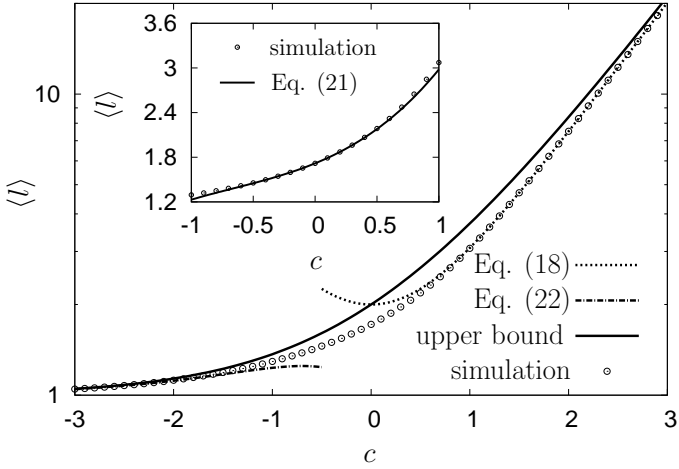}
\caption{\label{Fig:InfHl} Semi-logarithmic plot of the mean walk length
$\langle l \rangle$ as a function of the strength $c$ of the fitness gradient.
Simulation data are shown together with the approximations Eqs.~\eqref{Eq:cgg1},
\eqref{Eq:mcgg1}, and the upper bound Eq.~\eqref{Eq:upper}. Inset: Linear plot of $\langle l \rangle$ vs. $c$ together with
Eq.~\eqref{Eq:smallc}}
\end{figure}
In Fig.~\ref{Fig:InfHl}, we compare $\langle l \rangle$ obtained from simulations of $10^8$ independent
realization with sequence length $L=2^{30}$ to the approximations Eqs.~\eqref{Eq:cgg1},
\eqref{Eq:smallc}, and \eqref{Eq:mcgg1} together with the upper bound
of Eq.~\eqref{Eq:upper}.
The simulation method is explained in \ref{Sec:B}.
As a rule of thumb, the large $|c|$ approximations work well for $|c| > 1$ and
the approximation for $|c|\ll 1$ becomes accurate for $|c|<1$.

\subsubsection{The limit $c \to \infty$ at finite $L$}

\begin{figure}[t]
\includegraphics[width=\columnwidth]{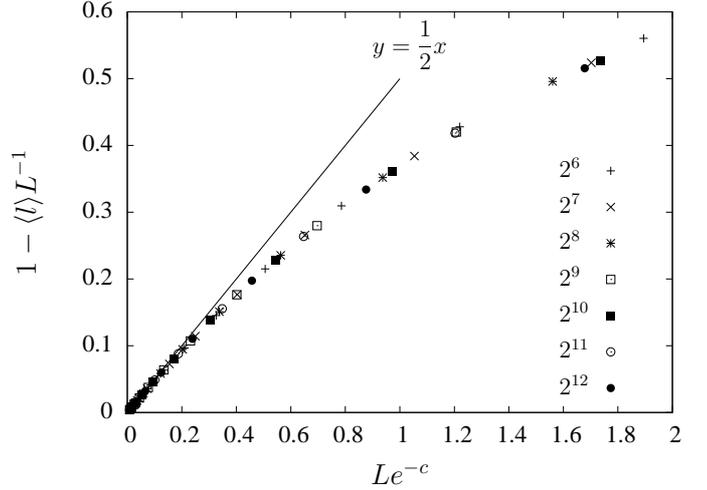}
\caption{\label{Fig:comp}  Mean walk length $\langle l \rangle$
is analyzed for large $c$ by
plotting $1 - \langle l \rangle L^{-1}$ vs. $L e^{-c}$ for $L=2^{6}$, $2^7,\ldots,$
$2^{12}$. All data points nicely collapse onto a single curve which is the scaling function
$\Lambda(x)$ in Eq.~\eqref{Eq:Lscale}. The asymptotic behavior $\Lambda(x) \approx
x/2$ for $x \to 0$ is also confirmed.
}
\end{figure}

For finite $L$, it is clear that the mean walk length should approach $L$ as $c\rightarrow \infty$.
This limit can be attained when $c$ is much larger than the (typical)
largest value among $L$ i.i.d. random variables. For the Gumbel case,
this corresponds to $\ln L \ll c$ or $L e^{-c} \ll 1$.
To find an approximate solution of $\langle l \rangle$ under this condition,
we go back to Eq.~\eqref{Eq:Hl} and expand $(L-k+1)/a_k$ in terms of $e^{-c}$ as
\begin{align}
\frac{L-k+1}{a_k}  
\approx 1 - e^{-c} \left ( 1 + \frac{1}{L-k+1} \right )
\end{align}
for $k\ge 2$ and $L/a_1 \approx 1- e^{-c}/L$,
where we have kept terms up to $O(e^{-c})$.
Hence
\begin{align}
H_l &\approx 1 - \frac{e^{-c}}{L}
- e^{-c} \sum_{k=2}^l\left ( 1 + \frac{1}{L-k+1} \right )\nonumber \\
&= 1 - e^{-c} (l-1) - e^{-c} \sum_{k=1}^l \frac{1}{L-k+1},
\end{align}
which gives
\begin{align}
\langle l \rangle 
&\approx L - e^{-c} \left ( \frac{L(L-1)}{2} + \sum_{l=1}^L \sum_{k=1}^l 
\frac{1}{L-k+1} \right ) \nonumber\\
&= L \left ( 1 - \frac{L+1}{2} e^{-c} \right ).
\end{align}
As anticipated, $L e^{-c}$ appears as an expansion parameter and $\langle l \rangle$ approches $L$
as $c\rightarrow \infty$. Thus, it is quite plausible to assume a scaling form such that
\begin{align}
1-\frac{\langle l \rangle}{L} = \Lambda\left ( L e^{-c} \right ),
\label{Eq:Lscale}
\end{align}
where $\Lambda(x)$ is a scaling function with asymptotic behavior
$\Lambda(x) \simeq x/2$ for sufficiently small $x$.
That is, if we plot $1-\langle l \rangle / L$ as a function of $L e^{-c}$ for sufficiently large $L$, 
the data obtained for different combinations of $L$ and $c$ should
collapse onto a single curve. To confirm this, we performed simulations for
$L$ ranging from $2^{6}$ to $2^{12}$. Figure \ref{Fig:comp} which is the result of $10^{8}$ independent
realizations for each data point indeed confirms the existence of such
a scaling
function.

\subsection{\label{Sec:arbid}Greedy walks with arbitrary starting point}
In this section, we relax the assumption that the walk always starts
at the antipodal sequence $\cC_a$ and calculate the mean number of steps in the
case that the initial genotype has Hamming distance $d_0$ from the reference
sequence $\cC_r$. Note that the case 
treated in the previous section correspond to $d_0=L$ 
and the case with $c<0$ in the previous section can be understood
as a greedy walk starting at $d_0=0$ with positive $c$. 
We consider the limit $L, d_0 \to \infty$ with
$\alpha = d_0/L$ kept finite. Since the RMF landscape is symmetric under the simultaneous transformations
$c \mapsto -c$ and $d_0 \mapsto L-d_0$, we can set $c$ to be non-negative without loss of generality.

\subsubsection{Exact asymptotic solution}

Unlike the previous section, the initial genotype has $O(L)$ neighbors
in both the uphill and downhill directions, and we cannot exclude the 
possibility that the walker takes a downhill step.
Assume that the walker arrives at the sequence $\cC_l$ at the
$l$-th step and $d_r(\cC_l) = d$.
Note that $d_0-d$ needs not be the same as $l$.
Now, we introduce the function  
\begin{equation}
\label{qdef}
q(y,\sigma) = \begin{cases}
\frac{d(F(y)^d)}{dy} \left ( F(y+ 2c ) \right )^{L-d} & \text{ if } \sigma  =+1,\\
\frac{d(F(y)^{L-d})}{dy} \left ( F(y- 2c ) \right )^{d} & \text{ if } \sigma  =-1,
\end{cases}
\end{equation}
which is interpreted as the probability density that the largest fitness 
among the uphill (downhill) neighbors has the random contribution $y$ and all downhill (uphill) neighbors
have smaller fitness when $\sigma = 1$ ($-1$).

As in \sref{Sec:antipodal}, the probability of taking at least $l$ steps is
denoted by $H_l$.
Since the walker may move in the uphill or downhill direction
with non-negligible probability, we have to take into account all
possible combinations of directions.
If $\sigma_l\in\{\pm 1\}$ is the change in the distance from the
antipodal sequence $\cC_a$ at the $l$-th step, then the
change in $d$ over a path is stored in an ordered set $\{\sigma\}_l = (\sigma_1,\sigma_2,\ldots, \sigma_l)$.
Defining $M_l \equiv \sum_{i=1}^l
\sigma_i$, the Hamming distance from $\cC_r$ after $l$ steps
is $d = d_0 - M_l$. We assume (and will subsequently verify) that the probability that the walker takes $O(L)$ steps is exponentially small
for large $L$. Accordingly, the scaled distance $d/L$ and therefore
the function $q(y,\sigma)$ in (\ref{qdef2}) do not change significantly
during the walk.  Within this assumption,
we can approximate (\ref{qdef}) in the form 
\begin{equation}
\label{qdef2}
q(y,\sigma) = L \beta s_\sigma Q(x+\sigma c),
\end{equation}
where $\beta = \alpha e^c + (1-\alpha)e^{-c}$, $s_1 = \alpha e^c/\beta$, $s_{-1} = 1 - s_1$,
and $Q(x) = \exp(-x-e^{-x} L \beta )$, which is independent of $l$.

Let $J_l(x,\{\sigma\}_l)$ be the probability that a walk has moved according to
$\{\sigma\}_l$ and 
the fitness of the sequence at the $l$th step is smaller
than $-c(d_0-M_l) + x$. With $j_l(x,\{\sigma\}_l) =
\frac{d}{dx}J_l(x,\{\sigma\}_l)$ we then have, in analogy to (\ref{Hldef}), 
\begin{align}
H_l = \sum_{\{\sigma\}_l} J_l (\infty,\{\sigma\}_l) = 
\sum_{\{\sigma\}_l} \int_{-\infty}^\infty j_l(x,\{\sigma\}_l) dx,
\label{Eq:HlJ}
\end{align}
where the summation is over all possible  $2^l$ combinations of $\{\sigma\}_l$.
Similar to (\ref{basicrec}) one can construct a recursion relation for
$j_l$, which reads
\begin{align}
j_l(x,\{\sigma\}_l) &= q(x,\sigma_l) J_{l-1}(x+\sigma_l c,\{\sigma\}_{l-1})\nonumber\\
&= q(x,\sigma_l) \int_{-\infty}^{x+\sigma_l c} j_{l-1}(y,\{\sigma\}_{l-1}) dy.
\label{Eq:jlsigma}
\end{align}
For $l=1$,  
\begin{align}
  \label{Eq:qsigma}
j_1(x,\{\sigma\}_1) &= q(x,\sigma_1) \int_{-\infty}^{x+\sigma_1 c} f(y) dy\nonumber\\
&= q(x,\sigma_1) F(x+\sigma_1 c) \approx q(x,\sigma_1),
\end{align}
where we have approximated $F \approx 1$ because the relevant fitness
values reside far in the tail of the distribution when $L$ is large.

If we neglect the effect of the change in $d$ on $q(x,\sigma_l)$ 
as assumed above, we get
\begin{align}
J_l(\infty,\{\sigma\}_l) &\approx (L \beta)^l
\left ( \prod_{k=1}^l s_{\sigma_k}\right )
\int_{-\infty}^\infty dy_l Q(y_l +\sigma_l c) \times \nonumber \\
&\times  {\prod_{k=l}^{2}} \hspace{-16pt}\left . \phantom{\prod} \right .' \int_{-\infty}^{y_{k} + \sigma_{k}c}
Q(y_{k-1} + \sigma_{k-1} c) dy_{k-1}  \nonumber
\\
&= 
\prod_{k=1}^l \frac{s_{\sigma_k}}{1 + \sum_{m=1}^{k-1} \exp(-c M_m)},
\label{Eq:JlS}
\end{align}
where $\prod '$ in the second line signifies an index-ordered product in 
descending order of $k$, which should be interpreted as 1 if $l=1$.
The solvability of the nested chain of integrals in (\ref{Eq:JlS}) is
specific to the Gumbel distribution; see \ref{Sec:A}. 
From Eqs.~\eqref{Eq:HlJ} and \eqref{Eq:JlS}, we arrive at our central result 
\begin{align}
H_l =\sum_{\{\sigma\}_l}
\prod_{k=1}^l \frac{s_{\sigma_k}}{1 + \sum_{m=1}^{k-1} \exp(-c M_m)} 
\label{Eq:Hld}
\end{align}
which reduces to Eq.~\eqref{Eq:HlLinf} when $\alpha = 1$.

\subsubsection{Dependence of the walk length on $\alpha$ and $c$}

Since
$\exp(-mc) \le \exp(-cM_m) \le \exp(mc)$
and $s_1+s_{-1} = 1$, the expression (\ref{Eq:Hld}) is bounded from
below and above by its values for $\alpha = 0$ and $\alpha = 1$,
respectively 
\begin{equation}
H_l|_{\alpha=0}=  \prod_{k=1}^l \frac{1-e^c}{1-e^{kc}}
\le H_l\le
\prod_{k=1}^l \frac{1-e^{-c}}{1-e^{-kc}} = H_l|_{\alpha=1}.
\end{equation}
In fact,  using $\frac{d}{d\alpha} s_1 = - \frac{d}{d\alpha} s_{-1} \ge 0$
and $ \exp(-c + A) \le \exp(c + A)$ for any real $A$, one can easily see
that 
$\frac{d}{d\alpha} H_l \ge 0$, where the equality holds only when $c=0$.
That is, $H_l$ is an increasing function of $\alpha$, and
correspondingly the mean walk length (\ref{Eq:deflength}) decreases monotonically as the
position of the starting point approaches the reference sequence,
which is easily conceivable. 

By contrast, the dependence of the mean walk length on $c$ is more
complex. We have seen above in Sec.~\ref{Sec:largesmallc} that the walk length
decreases with increasing $c$ when the walk starts at the reference
sequence ($\alpha = 0$), and we will now show that such an initial
decrease occurs whenever $\alpha < \frac{1}{2}$. On the other hand,
for very large $c$ the walk length must approach $\alpha L$ for any
$\alpha > 0$, and we must therefore expect a non-monotonic dependence
on $c$ for $0 < \alpha < \frac{1}{2}$. Such a behavior was already
reported by \cite{NSK2014} on the basis of numerical
simulations. 

When $c\ll 1$, we can approximate $H_l$ up to $O(c^2)$ as (see~\ref{Sec:A} for the derivation)
\begin{equation}
l! H_l = 1 + \delta\frac{\permu{l}2}{4} c + \delta^2 c^2 
\frac{9\permu{l}4 + 32 \permu{l}3}{288} + (1-\delta^2) c^2  \frac{7\permu{l}3}{108} ,
\label{Eq:Hlalpha}
\end{equation}
where $\delta = s_1 - s_{-1} = (\alpha e^c - (1-\alpha) e^{-c})/\beta$.
Accordingly, the mean number of steps becomes
\begin{align}
\langle l \rangle &\approx e - 1 + \frac{ \delta}{4} e c + 
\frac{41  \delta^2}{288} e c^2 + \frac{7 }{108} (1-\delta^2) e c^2\nonumber \\
&\approx e - 1 + \frac{2 \alpha - 1}{4} ec 
+ \frac{123 + 596 \alpha(1-\alpha)}{864} e c^2,
\label{Eq:lalpha}
\end{align}
where we have also expanded $\delta$ up to $O(c^2)$.
Hence $\langle l \rangle$ is an increasing function of $c$ for
$\alpha \ge \frac{1}{2} $ when $c$ is small enough, while for $\alpha < \frac{1}{2}$ the mean walk length initially decreases with $c$ for small $c$. Since the walk length is known to increase at large $c$, it follows that 
there must be a turning point which, in the quadratic approximation (\ref{Eq:lalpha}), is given by  
\begin{equation}
  \label{Eq:cturn}
c_\text{turn} \approx \frac{108 ( 1 - 2 \alpha)}{123 + 596 \alpha ( 1-\alpha)}.
\end{equation}

\begin{figure}[t]
\includegraphics[width=\columnwidth]{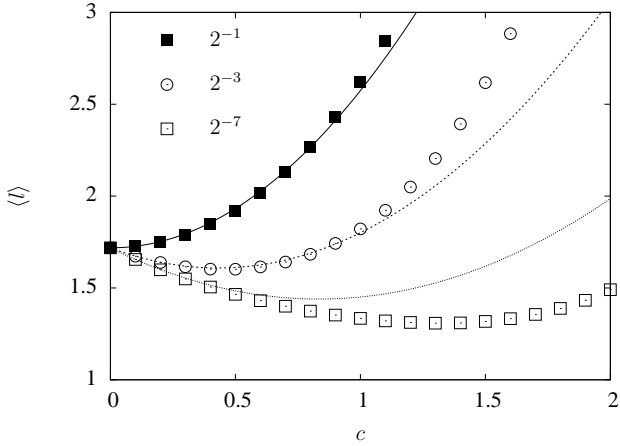}
\caption{\label{Fig:alpha}  The mean walk length $\langle l \rangle$
is plotted as a
function of $c$ for different starting points $\alpha = 2^{-1}$, $2^{-3}$, and
$2^{-7}$ (from top to bottom) with comparison to 
the expansion in Eq.~\eqref{Eq:lalpha}. The sequence length is 
$L=2^{60}$ and the number of independent runs for each data point is $10^9$.
}
\end{figure}
A comparison of Eq.~\eqref{Eq:lalpha}
with simulations is shown in
Fig.~\ref{Fig:alpha}, which illustrates the accuracy of the
analytic expression (\ref{Eq:lalpha}) for
small $c$. As predicted, it also confirms the absence of a turning point for
$\alpha\ge \frac{1}{2}$.
As $\alpha$ decreases, the position of the turning point found in the
simulations moves to larger $c$, which makes the small $c$ 
approximation inaccurate for precisely pinpointing $c_\text{turn}$. 

\begin{figure}[t]
\includegraphics[width=\columnwidth]{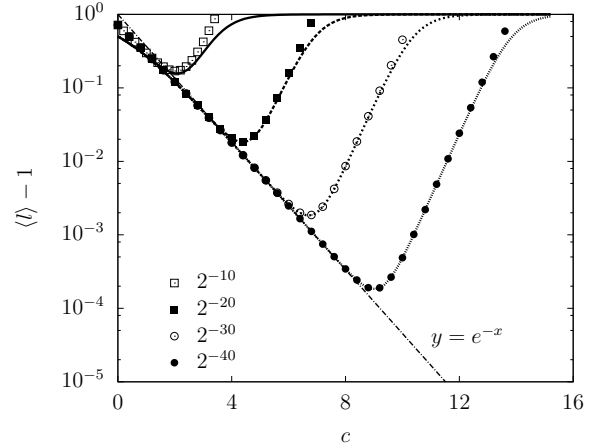}
\caption{\label{Fig:turn_small}  Semi-logarithmic plots of $\langle l \rangle - 1$
vs. $c$ for $\alpha = 2^{-10}$, $2^{-20}$, $2^{-30}$, and
$2^{-40}$ (from top to bottom) with comparison to Eq.~\eqref{Eq:H2alpha}. 
The sequence length is 
$L=2^{60}$ and the number of independent runs for each data
point is $10^9$.
}
\end{figure}

From Fig.~\ref{Fig:alpha}, the position $c_\text{turn}$ of the turning
point seems to diverge as $\alpha \to 0$. 
When $\alpha = 0$, the mean walk length decreases as $\langle l \rangle \approx 
1+ e^{-c}$ for sufficiently large $c$ as shown in \sref{Sec:largesmallc}.
When $\alpha$ is very small, $\langle l \rangle$ should therefore first decrease
as $1+ e^{-c}$, but eventually increase with $c$ for sufficiently large $c$.  
As in the case of $c<0$ and $|c|\gg
1$ in \sref{Sec:largesmallc}, when 
$\langle l \rangle -1 \ll 1$ this quantity is expected to be well approximated by
\begin{align}
\langle l \rangle -1\approx  
 H_2 = \frac{s_1}{1+e^{-c}} + \frac{s_{-1}}{1+e^c}
= \frac{1 + \varepsilon e^{3c}}{(1+e^c)(1+ \varepsilon e^{2 c})},
\label{Eq:H2alpha}
\end{align} 
where $\varepsilon = \alpha / (1 - \alpha)$.
In Fig.~\ref{Fig:turn_small}, we compare simulation results for small
$\alpha$ ($\alpha \le 2^{-10}$) with Eq.~\eqref{Eq:H2alpha}, which shows
an excellent agreement as long as $\langle l \rangle -1 \le 0.1$.
Hence the turning point can be found by investigating the minimum of $H_2$,
which gives  
\begin{equation}
\label{Eq:cturn2}
c_\text{turn} \approx -\frac{1}{3}\ln (2 \varepsilon),
\end{equation}
where we have only kept the leading order of $\varepsilon \ll 1$.
Note that $c_\text{turn}$ indeed diverges as $\alpha \rightarrow 0$.
When $c \le c_\text{turn}$, the mean walk length is well approximated
by $1 + e^{-c}$ which is the result for $\alpha = 0$ with $c \ge 1$.

To put these results into perspective and provide an intuitive
explanation of the observed non-monotonic behavior as a function of $c$, 
it is instructive to compare the mean walk length
to the density of local fitness maxima. Since the walk is trapped at local maxima,
one generally expects an inverse relationship between the two quantities
\citep{Weinberger1991,Nowak2015}. According to
\eqref{Eq:maxprob}, the density of local fitness maxima at distance
$d$ from the reference sequence becomes $p_c^{\mathrm{max}} \approx 1/(\beta L)$ in the limit when 
$L, d \to \infty$ at fixed $\alpha = d/L$, where we recall that 
$\beta = \alpha e^c + (1-\alpha) e^{-c}$. 
It is straightforward to check that $p_c^{\mathrm{max}}$ decreases monotonically
with increasing $\alpha$ but displays a maximum as a function of $c$
for $\alpha < \frac{1}{2}$. The maximum  is located at $\tilde{c}_{\text{turn}}
= - \frac{1}{2} \ln \varepsilon$ which is similar to
(\ref{Eq:cturn2}) and also diverges for $\alpha \to 0$. We may thus
conclude that, at least qualitatively, the behavior of the greedy walk
length reflects that of the density of local maxima. 

\section{\label{Sec:bound} General distribution of the random fitness
component}

\subsection{\label{Sec:genintro} Reformulation of the problem}

Up to now, we have presented a detailed analysis of greedy
adaptive walks for the case of Gumbel-distributed random fitness components.
In this section, we will generalize our findings to arbitrary
probability distribution functions $F(y)$, focusing on the limit $L
\rightarrow \infty$. 
As in \sref{Sec:arbid}, the initial genotype from which the walker starts has
the Hamming distance $d_0$ from the reference sequence and we take $d_0, L
\to \infty$ at fixed $\alpha = d_0/L$. Under these conditions the walker
takes both uphill and downhill steps. 
As long as the number of steps 
taken is much smaller than $L$, the walk dynamics can be formulated
in terms of the following game:

At each round $n$ ($n=1,2,\ldots$), one generates two random variables $Y_n$
and $Y_{-n}$, where $Y_n$ is drawn from the distribution $F(y)^{L\alpha}$
and $Y_{-n}$ from $F(y)^{L(1-\alpha)}$. Then choose the larger one 
between $Y_n + c$ and $Y_{-n}-c$. Assuming that the larger one
is $Y_{\sigma_n n} + \sigma_n c$ where $\sigma_n$ can be either 1 or $-1$, 
this number is compared to $X_{n-1}$, with $X_0 = -\infty$.
If $X_{n-1}$ is larger than $Y_{\sigma_n n} + \sigma_n c$ the game is over.
Otherwise, we set $X_n = Y_{\sigma_n n}$ and go to the next round.
Then the mean number of steps in the greedy walk is the same as the mean number of
rounds up to the end of the game.

For convenience, we introduce an event 
\begin{align}
E_n(\sigma) = \{Y_{\sigma  n}+
\sigma c > Y_{-\sigma n} - \sigma c ~\&~ 
Y_{\sigma  n}+ \sigma c > X_{n-1} \},
\end{align}
where $X_{n-1}$ is defined as above.
With this notation, we can write down the probability that the game 
persists at least up to $l$ rounds as
\begin{equation}
\label{Eq:Hlgen}
H_l = \sum_{\{\sigma\}_l}\text{Prob}(E_1(\sigma_1) \cap
E_2(\sigma_2) \cap \cdots \cap E_l(\sigma_l) ),
\end{equation}
where the summation is over all possible sequences of $\sigma$'s of
length $l$.

For $\alpha = 1$ all steps are in the uphill direction and
(\ref{Eq:Hlgen}) reduces to a single term with $\sigma_1 =
\sigma_2 = \cdots \sigma_l = 1$, which can be written as
\begin{equation}
\label{Eq:Hlalpha1}
H_l = \text{Prob}(Y_1 + c < Y_2 + 2c < \cdots < Y_l + cl),
\end{equation}
that is, the probability that the sequence of random variables $Y_n +
cn$ is ascendingly ordered. This quantity was studied by
\cite{Franke2010} who showed that it is given by 
(\ref{Eq:HlLinf}) when the $Y_n$'s are drawn from the Gumbel
distribution. To see why this result applies in the present context,
we note that the distribution function of the maximum among $L$ i.i.d. Gumbel
random variables is given by 
\begin{equation}
\label{Eq:Gumbelshift}
F(y)^L = \exp[-Le^{-y}] = \exp[-e^{-(y-\ln L)}] = F(y-\ln L),
\end{equation}
which is identical to the original distribution up to an overall
shift that doesn't affect the ordering probability
(\ref{Eq:Hlalpha1}).

\subsection{Extreme value classes}

In order to analyze the problem for general choices of the
distribution function $F(y)$, we exploit the fact that $F(y)^{L
  \alpha}$ and $F(y)^{L (1- \alpha)}$ converge to one of the extreme
value distributions when the limit $L \to \infty$ is combined with a
suitable rescaling of $y$ \citep{deHaan}. Specifically, we introduce random variables
$Z_k$ such that $Y_k = a_L Z_k + b_L$,
where $k$ is an integer, $a_L$ and $b_L$ are parameters that depend
on $L$ but not on $k$, and $a_L >0$.  The parameters $a_L$ 
and $b_L$ have to be chosen such that the distribution of $Z_k$ has a well defined limit
as $L \rightarrow \infty$, that is, such that
\begin{equation}
\label{Eq:Klimit}
K(z) = \lim_{L\rightarrow \infty} F(a_L z + b_L)^L
\end{equation}
exists and is non-degenerate.

In terms of the transformed random variables, the event $E_n(\sigma)$ can be recast as 
\begin{align}
E_n(\sigma) = \{Z_{\sigma  n}+
\sigma \tc > Z_{-\sigma n} - \sigma \tc~ \&~ 
Z_{\sigma  n}+ \sigma \tc > \tilde X_{n-1} \},
\label{Eq:game}
\end{align}
where $\tc = c/a_L$ and $\tilde X_{n} = Z_{\sigma_{n} n}$. 
In the following we apply this approach to the three classes of
extreme value distributions.

\paragraph{Gumbel class} As a representative of the Gumbel class of
extreme value theory we choose the Weibull distribution $F(y) = 1 -
e^{-y^\theta}$. Setting 
\begin{align}
Y_k = (\ln L)^{1/\theta} \left (1 + \frac{Z_k}{\theta \ln L } \right ) = (\ln L)^{1/\theta}
+ \frac{Z_k}{\theta (\ln L)^{1-1/\theta}},
\end{align}
the limit (\ref{Eq:Klimit}) becomes the Gumbel distribution
\begin{align}
\label{Eq:KGumbel}
K_G(z) =  e^{- e^{-z}}
\end{align}
with support $-\infty < z <\infty$, 
as can be seen using the approximation
$y^\theta  = \ln L ( 1 + z/(\theta \ln L))^\theta \approx \ln L + z + o(1/\ln L)$.
Accordingly, 
\begin{equation}
\label{Eq:ctilde}
\tc = \theta c (\ln L)^{1-1/\theta }.
\end{equation}
For the case of an exponential distribution ($\theta = 1$) it follows
that $\tc = c$, and we conclude that the results derived in
\sref{Sec:mean} for Gumbel-distributed random fitness components 
in fact apply asymptotically to \textit{all distributions with
exponential tails}.  On the other hand, when the tail of the
distribution is fatter ($\theta < 1$) or thinner ($\theta > 1$) than
exponential, $\tc$ asymptotically scales to zero or infinity,
respectively, when the limit $L \to \infty$ is taken at fixed $c$.
This implies that greedy adaptive walks on the RMF landscape behave
asymptotically like those on an uncorrelated landscape in the first
case, their length approaching $\langle l \rangle = e-1$, 
whereas in the second case the walks move all the way to the
reference sequence and $\langle l \rangle \to \alpha L$. Because of
the logarithmic dependence of $\tc$ on $L$, corrections to this
asymptotic behavior are however expected to be important, and can be
obtained from the results of \sref{Sec:mean} by replacing $c$ with $\tc$.  

\paragraph{Fr\'echet class} This class comprises distributions with a
power law tail and can be represented by $F(y) = 1 - y^{-\mu}$ with
$y>1$ and $\mu>0$. Choosing  
$a_L = L^{1/\mu}$ and $b_L =0$, the limit (\ref{Eq:Klimit}) becomes  
\begin{align}
\label{Eq:KFrechet}
K_F(z) = \lim_{L\rightarrow\infty} 
\left ( 1 - \frac{z^{-\mu}}{L} \right )^L = e^{- z^{-\mu}}
\end{align}
with the support $z > 0$. Accordingly, $\tc = c/L^{1/\mu}$.
Assuming that $c$ remains finite when taking the $L \rightarrow \infty$ limit,
$\tc$ approaches zero and the problem
becomes identical to the greedy walk on an uncorrelated landscape. 

\paragraph{Weibull class} Lastly, we consider distributions with
bounded support, as represented by the distribution function
$F(y) =1 - (1-y)^\nu$ with $y \in [0,1]$.
Setting $a_L = L^{-1/\nu}$ and $b_L = 1$, the limiting distribution is
\begin{align}
\label{Eq:KWeibull}
K_W(z) = e^{-(-z)^\nu}
\end{align}
with the support $z<0$. Hence, in this case $\tc = cL^{1/\nu}$.
For finite $c$, $\tc$ is effectively infinite so that $H_l = 1$
and $\langle l \rangle \approx \alpha L$. \\

To summarize the results of this section, we have shown that it is
only for distributions with exponential tails that the mean greedy
walk length displays a non-trivial dependence on $c$, and in this case
the results of  \sref{Sec:mean} carry over without modification. In
all other cases a non-trivial asymptotic behavior requires that the
strength of the fitness gradient $c$ is scaled with $L$ in such a way
that $\tc$ has a finite limit for $L \to \infty$. 

For the non-Gumbel
extreme value classes characterized by the limiting distributions
(\ref{Eq:KFrechet}) and (\ref{Eq:KWeibull}) a closed-form solution analogous    
to that obtained in \sref{Sec:mean} for the Gumbel class appears to be
out of reach, with the exception of the Weibull class with $\nu = 1$,
where the explicit formula
\begin{align}
\langle l\rangle = \left ( \sum_{k=0}^\infty \frac{(-1)^k}{k!} e^{-k(k-1)\tc /2}\right )^{-1}-1
\label{Eq:boundl}
\end{align}
can be derived for $\alpha = 1$  (see \ref{Sec:C}). In the general case we therefore resort to approximations that are valid
for small and large $\tc$, respectively. Apart from their intrinsic
interest, these results can be used to compute corrections to the
asymptotic walk length when $c$ and $L$ are both finite. Throughout we
assume a general limiting distribution function $K(z)$ with the
corresponding probability density $f_K(x) = \frac{dK}{dz}$.

\subsection{\label{Sec:smalltc} Small $\tc$ approximation}

We begin with the case $\alpha = 1$, where previous results for the ordering
probability (\ref{Eq:Hlalpha1}) can be exploited. Indeed, the results of
\cite{Franke2010} imply that 
\begin{equation}
\label{Eq:HlFranke}
H_l = \frac{1}{l!} + \frac{\tc}{(l-2)!} \int_{-\infty}^\infty dx \,
f_K(x)^2 + O(\tc^2)
\end{equation}
for $l \geq 2$ whenever the integral on the right hand side exists
(note that $H_1 = 1$ independent of $\tc$). Summing over $l$ it thus
follows that 
\begin{equation}
\label{Eq:lexp}
\langle l \rangle = e - 1 + e \tc \int_{-\infty}^\infty dx \,
f_K(x)^2 + O(\tc^2).
\end{equation}
Although the case of general $\alpha$ is more complex
[as can be seen by comparing the expressions (\ref{Eq:Hlgen}) and (\ref{Eq:Hlalpha1})],
the extensive calculations presented in \ref{Sec:app_small1} and \ref{Sec:app_small} yield a
simple result which amounts to replacing $\tc$ by $(2
\alpha - 1) \tc$ in (\ref{Eq:lexp}). Evaluating the integral over
$f_K(x)^2$ for the limiting distributions (\ref{Eq:KGumbel}) and (\ref{Eq:KFrechet}), we
thus obtain 
\begin{align}
\langle l \rangle_G &= e - 1 + \frac{2\alpha-1}{4} e\tc,\\
\langle l \rangle_F &= e - 1 +
(2\alpha-1) e \tc \mu 2^{-2-1/\mu}\Gamma\left (2+\frac{1}{\mu}\right )
\end{align}
to leading order in $\tc$. 
Note that the result for the Gumbel class is consistent with Eq.~\eqref{Eq:lalpha}.

\begin{figure}[t]
\includegraphics[width=\columnwidth]{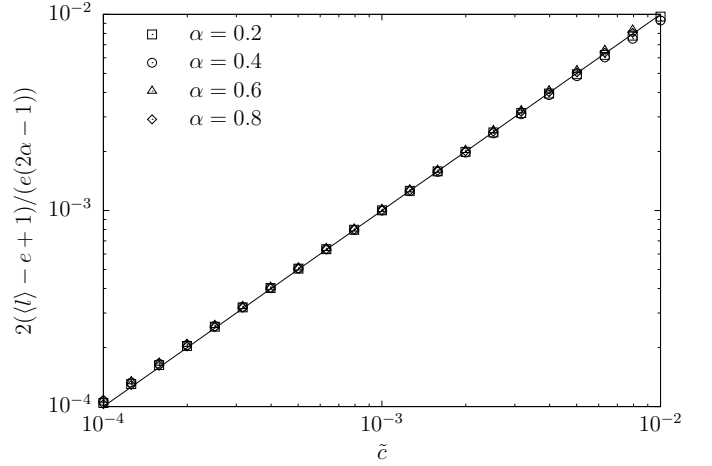}
\caption{\label{Fig:weibull_nu100} Double logarithmic plot of
$2(\langle l \rangle - e + 1)/(e (2\alpha - 1))$ vs. $\tc$ for the Weibull class
with $\nu = 1$ and various values of the scaled initial distance $\alpha$.
The data sets for different $\alpha$ collapse into the line $y=x$ as predicted by
Eq.~\eqref{Eq:wei_all}.}
\end{figure}
For the Weibull class, the integral on the right hand side of
(\ref{Eq:lexp}) exists only for $\nu > \frac{1}{2}$, and a more
careful analysis is required to find the leading correction in $\tc$.
Detailed calculations are found in \ref{Sec:app_small}. The final
result for the mean greedy walk length in this case reads,
\begin{align}
\frac{\langle l \rangle_W-e+1}{e(2\alpha-1)} = \begin{cases} 
\displaystyle  \nu 2^{-2+1/\nu} \Gamma\left (2-\frac{1}{\nu}\right ) \tc, & \nu > \frac{1}{2},\\
\\
\displaystyle -\frac{\tc}{4} \ln\left (e^{2\gamma-1} \tc\right ), & \nu = \frac{1}{2},\\
\\
\displaystyle  \frac{\Gamma(1-2\nu) \Gamma(\nu+1)}{2 \Gamma(1-\nu)} \tc^{2\nu},& \nu <\frac{1}{2}
\end{cases}
\label{Eq:wei_all}
\end{align}
where $\gamma\approx 0.5772$ is the Euler-Mascheroni constant.

\begin{figure}[t]
\includegraphics[width=\columnwidth]{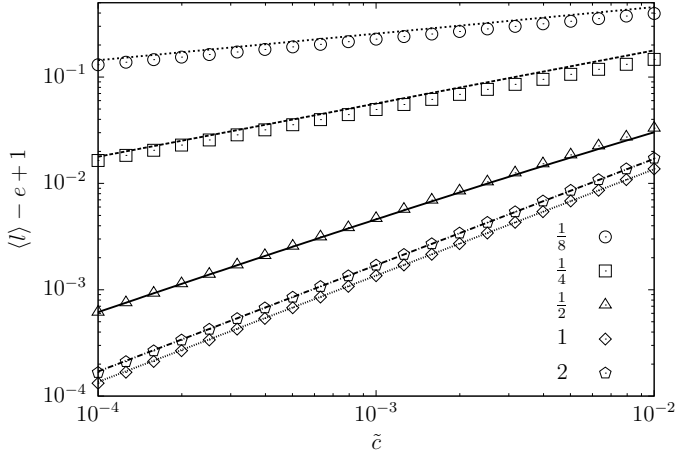}
\caption{\label{Fig:weibull_nu} Double logarithmic plot of
$\langle l \rangle - e + 1$ vs. $\tc$ for the Weibull class
with $\nu =2, 1$, $\frac{1}{2}$, $\frac{1}{4}$, and $\frac{1}{8}$. 
Here $\alpha$ is set to 1. 
The straight lines close to each data set 
are given by Eq.~\eqref{Eq:wei_all} with the corresponding values of $\nu$.
}
\end{figure}

In Fig.~\ref{Fig:weibull_nu100}, we confirm the validity of 
Eq.~\eqref{Eq:wei_all} for $K_W(x) = e^x$ ($x<0$, $\nu = 1$) by simulations.
Note that the result for $\nu = 1$ and $\alpha = 1$ can also be
obtained by expanding the exact expression (\ref{Eq:boundl}) up to $O(\tc)$.
In Fig.~\ref{Fig:weibull_nu} we show simulations of Eq.~\eqref{Eq:game}
for the Weibull class with various values of $\nu$ and $\alpha=1$. 
Each data symbol in Fig.~\ref{Fig:weibull_nu} is the result of $10^{11}$ 
independent runs.
The predicted Eq.~\eqref{Eq:wei_all} is in good agreement with the simulations.

\subsection{Large $\tc$ approximation}
When $\tc$ is very large, 
the walker takes uphill steps toward the reference state
with probability close to 1 as long as $\alpha \neq 0$. 
(If $\alpha = 0$ and $\tc$ is very large, 
$\langle l \rangle \approx 1 + H_2$ as for large negative $c$
in \sref{Sec:arbid}.)
In this case, we can neglect the effect
of downhill steps and the problem is reduced to the ordering problem studied
by \cite{Franke2010} with random variables drawn from a distribution $K(z)^\alpha$. 
If the support of $K(z)$ is unbounded from the above as in the
Gumbel and Fr\'echet classes, 
the walker stops at the $l$'th step when $Z_l$ happens to be larger than $\tc$.
Thus $\langle l\rangle $ can be estimated from the relation
$1 - K(\tc)^\alpha = 1/\langle l \rangle$.
For $K_G$, we get $\langle l \rangle \sim e^{\tc}/\alpha =\exp[\theta c (\ln L)^{1-1/\theta }]/\alpha$. 
For $K_F$, we get
\begin{align}
\langle l \rangle \sim \frac{\tc^{\mu}}{\alpha}
= \frac{c^\mu}{\alpha L}.
\end{align}

\begin{figure}[t]
\includegraphics[width=\columnwidth]{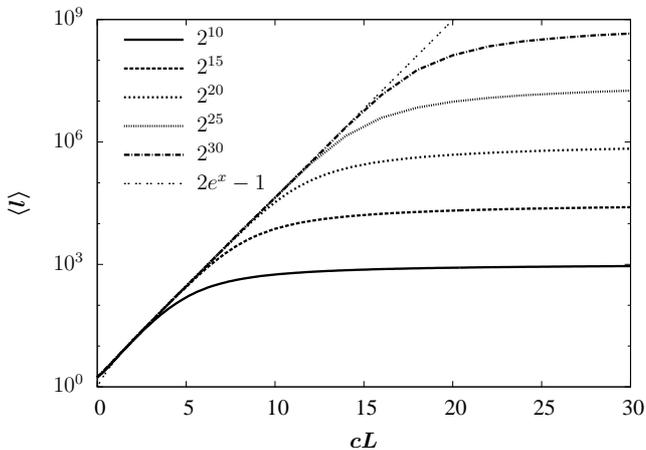}
\caption{\label{Fig:uni}  Mean walk length for uniformly distributed 
random fitness components with sequence length
$L = 2^{10}$, $2^{15}$, $2^{20}$, $2^{25}$, and $2^{30}$ (from bottom
to top) and antipodal starting point ($\alpha = 1$). 
As predicted by theory, in this case the walk length is a function
of $cL$. The number of runs is between
$10^{3}$ (for $L=2^{30}$) and $2 \times 10^4$ (for $L=2^{10}$). 
}
\end{figure}

If the support of $K(z)$ is bounded from above but
unbounded from below as in the Weibull class, the walker
should stop at the $(l-1)$th step when $Z_l$ happens to be smaller than 
$-\tc$. 
Thus $\langle l\rangle$ can be estimated from
$\langle l \rangle K(-\tc)^\alpha = 1$,
and using the expression for $K_W$, we get
$\langle l \rangle \sim e^{\alpha \tc^\nu} = e^{\alpha c^\nu L}$.
As an example, we present simulation results for the uniform distribution
[$F(x) = x$] with $\alpha=1$ in Fig.~\ref{Fig:uni}. Note that the leading 
behavior of Eq.~\eqref{Eq:boundl} for large $\tc$ is $2 e^{\tc} -1$,
which is consistent with the approximate estimate as well as with the 
simulation results in Fig.~\ref{Fig:uni}.

\section{\label{Sec:Discussion} Discussion and conclusion}

Adaptive walks arise as limiting cases from standard population genetic models and
represent an important paradigm in the theory of adaptation that has generated a number of 
non-trivial and experimentally testable predictions 
\citep{O2005,Schoustra2009,Seetharaman2014}. In particular, the greedy adaptive
walk considered in the present article is of biological interest for two reasons. First, it can be viewed 
as an approximate description of adaptation in a situation where the supply
of single beneficial mutations is high, such that all mutants are generated simultaneously and the mutation
of largest effect takes over by selection. Second, the greedy search strategy is arguably one that locates
local fitness maxima in the smallest possible number of steps. Greedy walks therefore provide important insights
into the geometry of high-dimensional random fitness landscapes, where, as shown by \cite{O2003}
for the uncorrelated case, fitness peaks are found within 2 mutational steps on average.

Here we have generalized the analysis of \cite{O2003} to the class of RMF models, where
a fitness gradient of strength $c$ is introduced to smoothen the landscape and to induce correlations between
genotypes. Fitness correlations are generally expected to increase the length of adaptive walks, and we show that this
is true in most but not all situations. 

Importantly, we find that the effect of the fitness gradient on the length of 
greedy walks depends crucially on the tail properties of the distribution underlying the random fitness component of
the RMF landscape, which can be classified in terms of extreme value theory (EVT). 
The results of our analysis in \sref{Sec:bound} imply that greedy walks on the RMF landscape
are asymptotically as short as in the uncorrelated case when the distribution of the random fitness contribution    
is heavy tailed (Fr\'echet or Gumbel with tail fatter than exponential) but become very long, with length equal
to the distance to the reference sequence, when the distribution is light tailed (Weibull or Gumbel with tail thinner
than exponential). Analogous results that single out fitness distributions with exponential tails with regard 
to structural properties of the RMF landscape (such as the number of 
local fitness peaks) and the length of random adaptive walks on this landscape have been reported previously
\citep{NSK2014,PSNK2014}. 

The prime representative of exponentially-tailed fitness distributions in the Gumbel class of EVT 
is the Gumbel distribution itself. For this case detailed results for the distribution of the greedy walk length
were obtained in \sref{Sec:mean}, for finite $L$ and antipodal starting point as well as for arbitrary starting points 
and $L \to \infty$. Perhaps the most surprising result of our analysis is the finding that the mean walk length
depends non-monotonically on the strength of the fitness gradient $c$, when the walk starts closer than at distance
$L/2$ from the reference sequence (i.e., the scaled distance is $\alpha < \frac{1}{2}$). 
This behavior was first
observed numerically by \cite{NSK2014}, and we have argued that it can be related to a similar non-monotonicity 
in the local density of fitness peaks. 

Although the analysis in \sref{Sec:arbid} is restricted to Gumbel-distributed
fitnesses, the fact that the leading order correction to the uncorrelated walk length derived in 
\sref{Sec:bound} is universally proportional to $2 \alpha -1$, and hence changes sign at $\alpha = \frac{1}{2}$,
indicates that the phenomenon is robust and does not depend on the distribution of the random fitness component.
Notably, within the EVT of adaptation it is usually assumed that the fitness of the wild type is high in absolute
terms \citep{G1984,O2002,O2005}. In the context of the RMF model this implies that the
adaptive walk starts rather close to the reference sequence, that is, at small $\alpha$, where the minimum in the
walk length as a function of $c$ is particularly pronounced (see \sref{Sec:arbid}).    

The existence of this minimum appears to contradict Orr's conjecture that
the $c=0$ value $\langle l \rangle = e-1$ constitutes a general lower bound  
on the length of adaptive walks \citep{O2003}. However, in formulating
his conjecture Orr demanded that the walk starts at a randomly chosen
point in sequence space, which implies that our result in Eq.~\eqref{Eq:lalpha} 
should be averaged over $\alpha$.
Since the  probability of choosing $\alpha$ is symmetric under the transformation
$\alpha \mapsto 1-\alpha$, the average of $2 \alpha - 1$ is zero and
that of $\alpha (1-\alpha)$ is positive. It follows that the averaged
walk length cannot be smaller than $e-1$. 
Similarly, in Rosenberg's refinement of Orr's conjecture
it is postulated that the fitness values in the landscape are
identically distributed, and that the fitness correlations between
neighboring genotypes are positive \citep[Sec. 5]{R2005}. Whereas the
latter statement applies to the RMF model \citep{NSK2014}, the former does
not. We conclude, therefore, that the seeming violation of the
conjecture of \cite{O2003} must be attributed
to the anisotropy of the RMF landscape.

An important aspect of adaptation that we have not addressed in this work concerns the fitness
level reached by the population at the end of the adaptive walk. In a recent comparative study
of different types of adaptive walks on Kauffman's NK-landscape, it was found that greedy walks 
reach higher fitness levels than random adaptive
walks on correlated landscapes, but the ranking among the walk types may change in the presence
of correlations \citep{Nowak2015}. The results presented here suggest that a detailed analysis
of the interplay between fitness correlations and the efficiency of different modes of adaptation
may be feasible within the framework of the RMF model, and we hope to report results along these lines
in the future.

\section*{Acknowledgments}
S-CP acknowledges the support by the Basic Science Research Program through the
National Research Foundation of Korea~(NRF) funded by the Ministry of
Science, ICT and Future Planning~(Grant No. 2014R1A1A2058694);
and by The Catholic University of Korea, Research Fund, 2015. JN and JK acknowledge support by Deutsche Forschungsgemeinschaft
within SFB-TR12, SPP 1590 and BCGS. Computations were performed
on the Cheops cluster at RRZK, Universit\"at zu K\"oln.

\appendix
\section{\label{Sec:B} Simulation Method}
Since we only need the largest value among a
certain number of i.i.d. random variables with a known distribution 
function, only two random numbers are necessary
to check if the walker can take a further step (see also \sref{Sec:genintro}). 
To be concrete, let us
assume that the walker is at $\cC$ with fitness $-d c + \xi$ where
$d$ is the Hamming distance of $\cC$ from the reference sequence.
Since the cumulative distribution of the largest random number among
$k$ variables is $F(x)^k$, $x$ can be generated by
$x = F^{-1}(y^{1/k})$, where $y$ is a uniformly distributed random
number ($0 < y < 1$).  For the Gumbel distribution, $x = \ln k - \ln (- \ln y)$ and for the uniform distribution, $x=y^{1/k}$.

If $x_1$ ($x_2$) is the largest random number among the uphill (downhill) neighbors
and if either $x_1+c$ or $x_2-c$ is larger than $\xi$, the walker
takes one further step. The actual direction will be determined
by checking whether $x_1+2c>x_2$ or not.
If $\xi$ is the largest, the walker stops. 

If $k$ is very large, we sometimes use the following
approximation
\begin{equation}
y^{1/k} = e^{\ln y /k} \approx 1 + \frac{\ln y}{k}
\left ( 1 + \frac{\ln y}{2 k} \right ).
\end{equation}
As a rule of thumb, for $k \ge 50~000$, the above approximation
gives a more accurate value than the direct power calculation
when we perform numerics with double precision ($\sim 10^{-15}$).
Note that when $k$ is very large,
it is better to use $1-x$ when deciding the fate of the walk, otherwise
there could be round-off errors which give $x=1$.

\section{\label{Sec:A} Derivation of Eq.~\eqref{Eq:Hlalpha}}
We first derive Eq.~\eqref{Eq:JlS}.
The integral over $y_1$ is readily calculated as
\begin{align}
L \beta \int_{-\infty}^{y_2 +\sigma_2 c} \exp(-y_1-\sigma_1 c - L\beta e^{-\sigma_1 c} e^{-y_1})
dy_{1} 
= \nonumber \\
\exp\left (-L \beta e^{-(\sigma_1 + \sigma_2) c} e^{-y_2} \right ).
\end{align}
Using the above equation, we can calculate the integral over $y_2$ as
\begin{align}
&L \beta \int_{-\infty}^{y_3 +\sigma_3 c} \exp(-y_2-\sigma_2 c - L\beta e^{-\sigma_2 c} (1+e^{-\sigma_1c} )e^{-y_2}  )
dy_{1} 
 \nonumber \\
&=\frac{1}{1+e^{-\sigma_1 c}}
\exp\left (-L \beta e^{-\sigma_3 c}\left (e^{-\sigma_1c} +  e^{-(\sigma_1 + \sigma_2) c} \right )e^{-y_3} \right ),
\end{align}
from which one can easily guess and prove that 
\begin{align}
(L\beta)^{n-1} {\prod_{k=n}^{2}} \hspace{-16pt}\left . \phantom{\prod} \right .'\int_{-\infty}^{y_k + \sigma_k c} 
Q(y_{k-1} + \sigma_{k-1} c ) d y_{k-1} \nonumber\\
= \prod_{k=1}^{n-1} \frac{1}{1 + \sum_{m=1}^{k-1} e^{-c M_m} } 
\exp\left (-L\beta e^{-\sigma_n c} \sum_{m=1}^{n-1} e^{-c M_m} e^{-y_n}\right ).
\end{align}
The final integral over $y_l$ gives Eq.~\eqref{Eq:JlS}.

For small $c$, we first expand the denominator in Eq.~\eqref{Eq:JlS} 
up to $O(c^2)$, which is
\begin{align}
1 + \sum_{m=1}^{k-1} e^{-c M_m}
= k \left (  1  - \frac{c}{k} \sum_{m=1}^{k-1} M_m + \frac{c^2}{2k}
\sum_{m=1}^{k-1} M_m ^2 \right ).
\end{align}
Then we expand the terms in $H_l$ up to $O(c^2)$ to get
\begin{align}
l! H_l = \sum_{\{\sigma\}} S(\{\sigma\}) 
\left ( 1 + c \gamma_1 + \frac{c^2}{2} \left ( \gamma_2 + \gamma_3 - \gamma_4 \right ) \right ),
\label{Eq:smallcd}
\end{align}
where $S(\{\sigma \} ) = \prod_{k=1}^l s_{\sigma_k}$ and
\begin{align}
\gamma_1 &=\sum_{k=1}^{l}\frac{1}{k}    
 \sum_{m=1}^{k-1} \sum_{n=1}^m \sigma_n ,\\
\gamma_2 &= \sum_{k=1}^l \frac{1}{k^2}\left (\sum_{m=1}^{k-1} \sum_{n=1}^m \sigma_n \right )^2  ,\\
\gamma_3 &= \left ( \sum_{k=1}^{l}   \frac{1}{k}
 \sum_{m=1}^{k-1} \sum_{n=1}^m \sigma_n \right )^2,\\
\gamma_4 &= \sum_{k=1}^l \frac{1}{k}\sum_{m=1}^{k-1}\left ( \sum_{n=1}^m \sigma_n \right )^2 .
\end{align}
The summations over $\sigma$ in Eq.~\eqref{Eq:smallcd} have two forms
\begin{align}
\sum_{\{\sigma\}}S(\{\sigma\}) \sigma_m &= (s_1 - s_{-1}) 
\prod_{k\neq m}\sum_{\sigma_k} s_{\sigma_k} =\delta,\\
\sum_{\{\sigma\}}S(\{\sigma\}) \sigma_m \sigma_n 
&=\delta^2 + \delta_{mn} (1-\delta^2)
\end{align} 
where $\delta = s_1 - s_{-1}$ and we have used $s_1 + s_{-1} = 1$ and $\sigma_n^2 = 1$.
Thus, we get
\begin{align}
\gamma_1 
&=\sum_{k=1}^l \frac{1}{k} \sum_{m=1}^{k-1} \sum_{n=1}^m \sum_{\sigma} S(\{\sigma\}) \sigma_n
= \delta \sum_{k=1}^l \frac{1}{k} \sum_{m=1}^{k-1} m \nonumber \\
&= \frac{l (l-1)}{4} \delta,
\end{align}
\begin{align}
\gamma_2 &= \sum_{k=1}^l \frac{1}{k^2}
\sum_{m=1}^{k-1} \sum_{n=1}^m
\sum_{r=1}^{k-1} \sum_{s=1}^r \sum_{\{\sigma\}}S(\{\sigma\}) \sigma_s \sigma_n\nonumber \\
&= \sum_{k=1}^l \frac{1}{k^2}
\sum_{m=1}^{k-1} \sum_{n=1}^m
\sum_{r=1}^{k-1} \sum_{s=1}^r \left ( \delta^2 + \delta_{sn} ( 1 - \delta^2)
\right )\nonumber \\
&=
\delta^2 \sum_{k=1}^l \frac{(k-1)^2}{4} 
+ (1-\delta^2) \sum_{k=1}^l \frac{1}{k^2}
\sum_{m=1}^{k-1} \sum_{r=1}^{k-1} \text{min}(m,r)\nonumber \\
&= \delta^2 \frac{l(l-1)(2l-1)}{24}
+ (1-\delta^2) \sum_{k=1}^l 
\frac{(k-1)(2k-1)}{6k}\nonumber \\
&=
\frac{\delta^2}{24} l (l-1)(2l-1)
+  \frac{1-\delta^2}{6}
\left ( l^2 - 2 l + \text{Har}[l] \right ),
\end{align}
where $\text{Har}[l] = \sum_{k=1}^l k^{-1}$,
and 
\begin{align}
\gamma_3 &= \sum_{k_1=1}^l \sum_{k_2=1}^l \frac{1}{k_1k_2} \sum_{m=1}^{k_1-1} \sum_{n=1}^m \sum_{r=1}^{k_2-1} \sum_{s=1}^r \left ( \delta^2 + \delta_{sn} ( 1- \delta^2)
\right )\nonumber \\
&=\delta^2 \left ( \sum_{k=1}^l \frac{k-1}{2} \right )^2 
+ (1-\delta^2) \sum_{k_1=1}^l \sum_{k_2=1}^l \frac{1}{k_1k_2} \sum_{m=1}^{k_1-1}\sum_{r=1}^{k_2-1} \text{min}(m,r)\nonumber \\
&=\delta^2 \frac{(l(l-1))^2}{16} + (1-\delta^2)
\sum_{k_1=1}^l \sum_{k_2=1}^l \frac{(k-1)(3K - k - 1)}{6 K},
\end{align}
where $K=\text{max}(k_1,k_2)$ and $k = \text{min}(k_1,k_2)$.
The summation in the last line of the above equation is
\begin{align}
&\sum_{k=1}^l \frac{(k-1)(2k-1)}{6k} + 2 \sum_{K=2}^l \sum_{k=1}^{K-1}
\frac{(k-1)(3K-k-1)}{6K}\nonumber\\
&=\frac{l(14 l^2 - 33 l + 37)}{108} - \frac{1}{6} \text{Har}[l].
\end{align}
And finally
\begin{align}
\gamma_4 &= \sum_{k=1}^l \frac{1}{k} \sum_{m=1}^{k-1} \sum_{n=1}^m \sum_{r=1}^m
\sum_{\{\sigma\}} S(\{\sigma\}) \sigma_n \sigma_r \nonumber \\
&= \sum_{k=1}^l \frac{1}{k} \sum_{m=1}^{k-1} \sum_{n=1}^m \sum_{r=1}^m
\left ( \delta^2 + \delta_{nr}(1-\delta^2) \right ) \nonumber \\
&= \sum_{k=1}^l \frac{1}{k} \sum_{m=1}^{k-1} 
\left ( \delta^2 m^2 + m (1-\delta^2) \right )\nonumber \\
&= \frac{l(l-1)}{4} + \delta^2\frac{l(l-1)(l-2)}{9} 
\end{align}
Combining these results, we arrive at Eq.~\eqref{Eq:Hlalpha}.

\section{\label{Sec:C}Derivation of Eq.~\eqref{Eq:boundl}}
This appendix calculates the mean
walk length for the case of $K(x) = e^{x}$ ($x<0$) with $\alpha = 1$.
For brevity, we will denote the mean walk length by $\ell$ and we drop
the tilde in $\tc$ in this appendix.
For completeness, we write the probability $H_l$ of taking at least $l$ steps 
\begin{align}
H_l = \int_{-\infty}^0 j_l(x) dy,
\end{align}
where $j_l(x)$ satisfies the recursion relation
\begin{align}
j_l(x) = e^{x} \int_{-\infty}^{x+c} j_{l-1}(x),
\label{Eq:jre_x}
\end{align}
with $j_1(x) = e^x$.
Since the support of $j_l(x)$ is $x<0$, Eq.~\eqref{Eq:jre_x} for $-c<x$ should be 
interpreted as 
\begin{align}
j_l(x) = e^x H_{l-1}.
\end{align}
Let 
\begin{align}
\varphi(x) = \int_{-\infty}^x \sum_{l=1}^\infty j_l(y) dy.
\end{align}
which is related to the mean walk distance by $\ell = \sum_l H_l =\varphi(0)$.
By summing both sides of 
Eq.~\eqref{Eq:jre_x} from $l=2$ to infinity, we get the difference-differential equation
\begin{align}
\frac{d\varphi(x)}{dx} = e^x ( 1 + \varphi(x+c) ),
\label{Eq:varphirecur}
\end{align}
where $\varphi(x)$ with $x>0$ should be interpreted as $\ell$.
Thus, for $x>-c$, we have
\begin{align}
\varphi(x) - \varphi(-c) = (1 + \ell ) (e^x - e^{-c} ).
\end{align}
Using $\varphi(0) = \ell$, we get $\varphi(-c) = (1+\ell) e^{-c}-1$ which,
in turn, gives, for $x>-c$,
\begin{align}
\varphi(x) = (1+\ell) e^x - 1.
\end{align}
Having determined $\varphi(x)$ for $x>-c$, we can find $\varphi(x)$ for $-2c<x<-c$ 
and so on.
After a few attempts, we make an ansatz,
for $-nc < x < -(n-1)c$,
\begin{align}
\varphi(x) = (1+\ell) \sum_{k=0}^n 
\frac{a_{n-k}}{k!}  \exp(kx+ k(k-1)c/2) -1,
\label{Eq:varphin}
\end{align}
which satisfies Eq.~\eqref{Eq:varphirecur}.
From the continuity of $\varphi$ at $x = -nc$, that is,
 $\varphi(-nc+0) = \varphi(-nc-0)$, we get a recursion
relation for the $a_n$ as
\begin{align}
a_{n+1} = e^{-n(n+1)c/2} \sum_{k=0}^n a_k \left ( \frac{e^{k(k+1)c/2}}{(n-k)!}
- \frac{e^{k(k-1)c/2}}{(n-k+1)!} \right ),
\end{align}
which is identical to
\begin{align}
\left ( a_{n+1} - a_n \right )e^{n(n+1)c/2}
= &-\sum_{k=0}^{n-1} \frac{1}{(n-k)!} \left ( a_{k+1} - a_k \right ) e^{k(k+1)c/2} \nonumber \\&- \frac{1}{(n+1)!},
\end{align}
with $a_0=1$ and $a_1=0$.
If we define $d_k = (k+1)!(a_{k+1} - a_k) e^{k(k+1) c/2}$, we get
\begin{align}
d_n = -\sum_{k=0}^{n-1}\binom{n+1}{k+1}d_k - 1
\end{align}
or
\begin{align}
\sum_{k=0}^n \binom{n+1}{k+1} d_{k} = -1.
\end{align}
Since $d_0=-1$ and $\sum_{k=0}^n \binom{n+1}{k+1} (-1)^{k+1} = -1$,
we conclude that $d_k = (-1)^{k+1}$.
That is, we get the recursion
\begin{align}
a_{n+1} -a_n = \frac{(-1)^{n+1}}{(n+1)!}e^{-n(n+1)c/2}
\end{align}
which is solved by 
\begin{align}
a_n = \sum_{k=0}^{n} \frac{(-1)^k}{k!} e^{-k(k-1)c/2}.
\end{align}

The mean walk length $\ell$ is determined by the boundary condition $\varphi(-\infty) = 0$.
Since $e^{kx}$ decays exponentially to zero unless $k=0$, 
this condition becomes
\begin{align}
0 =\varphi(-\infty) = -1 + (\ell + 1) \lim_{n\rightarrow \infty} a_n 
\end{align}
which gives Eq.~\eqref{Eq:boundl}.

\section{\label{Sec:app_small1} Small $\tc$ behavior : formal derivation}
In this appendix, we present a formal derivation of the
small $\tc$ behavior reported in \sref{Sec:smalltc}.  
In analogy to \eqref{qdef} we first introduce
\begin{align}
q_K(z,\sigma,\tc) = \omega_\sigma f_K(z) \left [ \frac{K(z+2\sigma\tc)}{K(z)} \right ]^{1-\omega_\sigma},
\end{align}
where $\omega_{+1}= \alpha$, $\omega_{-1}=1-\alpha$, and $f_K(z) = \frac{dK(z)}{dz}$.
We also  introduce $j_l$ iteratively as
\begin{align}
j_l(x,\{\sigma\}_l) 
&= q_K(x,\sigma_l,\tc) \int_{-\infty}^{x+\sigma_l \tc} j_{l-1}(y,\{\sigma\}_{l-1}) dy,
\label{Eq:jrecur}
\end{align}
with $j_1(x,\{\sigma\}_1) = q_K(x,\sigma_1,\tc)$.
Using $j_l$, we can write
\begin{align}
\langle l \rangle = 
\sum_{l=1}^\infty \sum_{\{\sigma\}_l} \int_{-\infty}^\infty j_l(x,\{\sigma\}_l) dx,
\label{Eq:KHlJ}
\end{align}
where $\sum_{\{\sigma\}_l}$ stands for the summation over all possible
$\sigma_i$'s for $i=1,\ldots,l$.

Since the mean walk distance for $\tc=0$ is $e-1$ for any distribution, 
$\langle l \rangle$ should take the form
\begin{align}
\langle l \rangle = e-1 + \lambda(\tc)
\label{Eq:def_lambda}
\end{align} 
with the property that $\lambda(\tc) \rightarrow 0$ 
as $\tc \rightarrow 0$.
In this appendix, we find the leading behavior of $\lambda(\tc)$ for small
$\tc\ll 1$.  At first, 
we decompose $j_l(x,\{\sigma\}_l) = j_l^{(0)}(x,\{\sigma\}_l)
+ g_l(x,\{\sigma\}_l,\tc)$,  where
\begin{align}
j_l^{(0)}\left (x,\{\sigma\}_l\right) 
&= q_K^0(x,\sigma_l) \frac{\A{l-1}}{(l-1)!} K(x)^{l-1}
\nonumber \\
&= \A{l} \frac{1}{l!} \frac{d}{dx} \left [ K(x) \right ]^l ,
\end{align} 
with 
\begin{align}
q_K^{0}(z,\sigma) &\equiv q_K(z,\sigma,\tc=0) = \omega_\sigma f_K(z), \\
\A{l} &\equiv \prod_{i=1}^l \omega_{\sigma_i},
\quad \A{0} \equiv 1,
\end{align}
and $g_l$ satisfies the recursion relation
\begin{align}
g_{l}(x,\{\sigma\}_{l},\tc) =& 
k_{l,0}(x,\{\sigma\}_{l},\tc)
\nonumber \\
&+ q_K(x,\sigma_{l},\tc)\int_{-\infty}^{x+\sigma_{l}\tc} g_{l-1}(y,\{\sigma\}_{l-1},\tc) dy,
\end{align}
with
\begin{align}
k_{l,0}(x,\{\sigma\}_{l},\tc) \equiv \frac{\A{l-1} }{(l-1)!}
\Biggl [ &q_K(x,\sigma_l,\tc) K(x+\sigma_{l}\tc)^{l-1} \nonumber \\
&- q_K^{0}(x,\sigma_l) K(x)^{l-1} \Biggr ].
\end{align}
Note that $j_l^{0}$ is the solution of Eq.~\eqref{Eq:jrecur} for $\tc = 0$.
Defining $k_{l,m}$ recursively as ($m \ge 1$)
\begin{align}
\frac{k_{l,m}(x,\{\sigma\}_{l+m},\tc)}{q_K(x,\sigma_{l+m},\tc)}
=\int_{-\infty}^{x+\sigma_{l+m}\tc} k_{l,m-1}(y,\{\sigma\}_{l+m-1},\tc) dy,
\end{align}
we can formally write 
\begin{align}
g_{l}\left (x,\{\sigma\}_l,\tc \right ) = \sum_{m=0}^{l-1}
k_{l-m,m}\left (x,\{\sigma\}_l,\tc \right ).
\end{align}
If we define 
\begin{align}
a_{l,m}\left (\{\sigma\}_{l+m},\tc \right ) \equiv \int_{-\infty}^\infty k_{l,m}\left (x,\{\sigma\}_{l+m},\tc \right ) dx ,
\end{align}
we can write
\begin{align}
\lambda(\tc) = \sum_{l=1}^\infty \sum_{m=0}^\infty \sum_{\{\sigma\}} 
a_{l,m}\left (\{\sigma\}_{l+m},\tc\right ) ,
\end{align}
where $\lambda(\tc)$ is defined in Eq.~\eqref{Eq:def_lambda}.
Since, for any $x$ and $\sigma$,
\begin{align}
\left |\int_{-\infty}^{x+\sigma \tc} k_{l,m}(y,\{\sigma\}_{l+m}, \tc) dy \right | 
&\le \left |\int_{-\infty}^{\infty} k_{l,m}(y,\{\sigma\}_{l+m}, \tc) dy \right| ,
\end{align}
we get an inequality for any $l$ and $m$ such as
\begin{align}
\left |a_{l,m} \right | 
\le \int_{-\infty}^\infty q_K(x,\sigma_{l+m},\tc) dx \left | a_{l,m-1} \right | 
= |a_{l,m-1}|
\le \left | a_{l,0} \right |,
\end{align}
which shows that $a_{l,m} = O\left (a_{l,0}\right )$. 

To extract the leading behavior of $\lambda(\tc)$, we consider the derivative
of $a_{l,m}$ with respect to $\tc$. 
That is, we consider
\begin{align}
b_{l,m}\left (\{\sigma\}_{l+m},\tc \right ) &
\equiv \frac{\partial}{\partial \tc} a_{l,m} \left (\{\sigma\}_{l+m},\tc \right )
\nonumber \\
&= \int_{-\infty}^\infty \tilde k_{l,m}(x,\{\sigma\}_{l+m},\tc) dx,
\end{align}
where $\tilde k_{l,m}$ is defined as
\begin{align}
\tilde k_{l,m}(x,\{\sigma\}_{l+m},\tc) \equiv \frac{\partial}{\partial \tc}
k_{l,m}(x,\{\sigma\}_{l+m},\tc).
\end{align}
Notice that
\begin{align}
\frac{d\lambda(\tc)}{d\tc} = \sum_{l=1}^\infty \sum_{m=0}^\infty \sum_{\{\sigma\}} 
b_{l,m}\left (\{\sigma\}_{l+m},\tc\right ) .
\end{align}

If $b_{l,m} = O(\tc^\eta)$ with $\eta \le 1$ for all $m$, we can write $\tilde k_{l,m}$
as the sum of $\kappa_{l,m}$ and $R_{l,m}$ which have the property that
\begin{align}
\int_{-\infty}^\infty \kappa_{l,m}(y,\{\sigma\}_{l+m},\tc) dy = O(\tc^\eta),\nonumber\\
\int_{-\infty}^\infty R_{l,m}(y,\{\sigma\}_{l+m},\tc) dy = o(\tc^\eta).
\end{align}

Now we will find a recursion relation for $\kappa_{l,m}$ from the exact relation,
$\tilde k_{l,m}(x,\{\sigma\}_{l+m},\tc) = \sum_{i=1}^6 I_i$,
where
\begin{align*}
I_1 &= q_K^{0}(x,\sigma_{l+m})\int_{-\infty}^x \kappa_{l,m-1}(y,\{\sigma\}_{l+m-1},\tc) dy,\\
I_2 &= q_K^{0}(x,\sigma_{l+m})\int_{-\infty}^x R_{l,m-1}(y,\{\sigma\}_{l+m-1},\tc) dy,\\
I_3 &= q_K^{0}(x,\sigma_{l+m})\int_{x}^{x+\sigma_{l+m}\tc} \tilde k_{l,m-1}(y,\{\sigma\}_{l+m-1},\tc) dy \\
I_4&=q_K^{1}(x,\sigma_{l+m},\tc) \int_{-\infty}^{x+\sigma_{l+m}\tc} k_{l,m-1}(y,\{\sigma\}_{l+m-1},\tc) dy,\\
I_5& = q_K(x,\sigma_{l+m},\tc) k_{l,m-1}(x+\sigma_{l+m} \tc,\{\sigma\}_{l+m-1},\tc),\\
I_6 &= \left [q_K-q_K^{0} \right ] \int_{-\infty}^{x+\sigma_{l+m}\tc} \tilde k_{l,m-1}(y,\{\sigma\}_{l+m-1},\tc) dy,
\end{align*}
and
\begin{align}
&q_K^{1}(x,\sigma,\tc) = \frac{\partial}{\partial \tc} q_K(x,\sigma,\tc) \nonumber \\
&= 2 \sigma \alpha (1-\alpha)\frac{ f_K(x) f_K(x+2\sigma \tc) }{K(x+2\sigma \tc)}
\left ( \frac{K(x+2 \sigma \tc)}{K(x)} \right )^{1-\omega_\sigma}.
\end{align}
Since $I_3$, $I_4$, $I_5$, $I_6$ are zero if $\tc = 0$ (note that $k_{l,m}$ 
is zero if $\tc=0$), the
integrals over these functions are $o(1)$ and they contribute to the remainder terms $R_{l,m}$.
Since the integral of $I_2$ should be $o({\tc}^\eta)$, 
we find $R_{l,m} = \sum_{i=2}^5 I_i$ and
\begin{align}
\kappa_{l,m}(x,\{\sigma\}_{l+m},\tc) = q_K^{0}(x,\sigma_{l+m}) \int_{-\infty}^x \kappa_{l,m-1}(y,\{\sigma\}_{l+m-1},\tc) dy.
\end{align}
In fact, the above consideration reveals that if we choose $R_{l,0}$ such
that the integral of $R_{l,0}(y)$ is $o(1)$, the
analysis of $\kappa_{l,m}$ should give all terms up to $O(1)$.

If we define 
\begin{align}
\phi(x) = \sum_{l=1}^\infty \sum_{m=0}^\infty \sum_{\{\sigma\}_{l+m}}\int_{-\infty}^x \kappa_{l,m}(y,\{\sigma\}_{l+m},\tc) dy,
\end{align}
we can write a differential equation for $\phi(x)$ such as
\begin{align}
\frac{d\phi(x)}{dx} = f_K(x) \phi(x) + \chi(x,\tc)
\label{Eq:phi}
\end{align}
where
\begin{align}
\chi(x,\tc) = \sum_{l=1}^\infty \sum_{\{\sigma\}_l} \kappa_{l,0}(x,\{\sigma\}_l,\tc),
\end{align}
and we have used $\sum_\sigma q_K^{0}(x,\sigma) = f_K(x)$.
The solution of Eq.~\eqref{Eq:phi} is 
\begin{align}
\phi(x) = e^{K(x)}\int_{-\infty}^x e^{-K(y)} \chi(y,\tc) dy,
\end{align}
which is related to $\lambda(\tc)$ as
\begin{align}
\frac{d\lambda(\tc)}{d\tc} 
\approx \lim_{x\rightarrow \infty}\phi(x) = e \int_{-\infty}^\infty
e^{-K(y)}\chi(y,\tc) dy = e \psi(\tc),
\label{Eq:ltc}
\end{align}
and, in turn,
\begin{align}
\langle l \rangle \approx e - 1 + e \int_0^\tc \psi(x) dx,
\end{align}
where the definition of $\psi$ is clear from the context.

If we choose $\kappa_{l,0} = \tilde k_{l,0}$ and $R_{l,0} = 0$, 
$\phi(x)$ can be used to find all terms up to $O(1)$.
With this choice, we get
\begin{align}
\chi&(x,\tc) = \frac{\partial}{\partial \tc} \left [ \sum_{l=1}^\infty 
\sum_{\{\sigma\}_{l}}k_{l,0}(x,\{\sigma\}_l,\tc) \right ] \\
&= \sum_{\sigma}\left [ q_K^{1}(x,\sigma,\tc) 
+ \sigma q_K(x,\sigma,\tc) f_K(x+\sigma\tc)\right ] e^{K(x+\sigma\tc)}.\nonumber
\end{align}

\section{\label{Sec:app_small} Small $\tc$ behavior : explict formulae
}
This appendix is a continuation of \ref{Sec:app_small1} and presents  
the explicit small $\tc$ behavior for the various classes. 
We first assume that $\psi(0)$ is non-zero. If this is true,
the mean walk distance becomes
\begin{align}
\langle l \rangle = e-1+e \tc \psi(0) + o(\tc).
\label{Eq:linear_l}
\end{align}
Since $\sum_\sigma q_K^1(x,\sigma,\tc=0) = 0$ and, in turn,
$\chi(x,0) = (2\alpha-1) f_K(x)^2 \exp[K(x)]$,
we get
\begin{align}
\psi(0) = e (2\alpha-1) \int_{-\infty}^\infty f_K(x)^2 dx,
\end{align}
and, accordingly,
\begin{align}
\langle l \rangle = e-1 + e\tc (2\alpha-1) \int_{-\infty}^\infty f_K(x)^2 + o(\tc),
\end{align}
as long as the integral 
is finite. As advertised, this generalizes Eq.~\eqref{Eq:lexp} to $\alpha < 1$. 
For the Gumbel and Fr\'echet classes and for 
the Weibull class with $\nu > \frac{1}{2}$, the integral becomes
\begin{align}
  \int_{-\infty}^\infty e^{-2x - 2 e^{-x}} dx &= \frac{1}{4},\\
 \int_0^\infty \mu^2 x^{-2(\mu+1)} e^{-2x^{-\mu}} dx &= 
\mu 2^{-2-1/\mu}\Gamma\left (2+\mu^{-1}\right ),\\
\int_{-\infty}^0 \nu^2 (-x)^{2(\nu-1)} e^{-2 (-x)^\nu} dx
&= \nu 2^{-2+1/\nu}\Gamma\left ( 2 - \nu^{-1}\right ).
\label{Eq:weibull_small}
\end{align}

For the Weibull class with $\nu \le \frac{1}{2}$, Eq.~\eqref{Eq:weibull_small}
is not applicable and we have to be more
careful to find the small $\tc$ behavior of $\psi(\tc)$ for this case. 
To this end, we first write
\begin{align}
\psi(\tc)
= \sum_{\sigma} \left [N_1(\sigma,\tc) + N_2(\sigma,\tc) \right ],
\label{Eq:Wei_half}
\end{align}
where 
\begin{align}
\label{Eq:defN1}
N_1(\sigma,\tc) &= \int_{-\infty}^\infty q_K^{1}(x,\sigma,\tc) e^{K(x+\sigma \tc) - K(x)} dx,\\
N_2(\sigma,\tc) &= \sigma \int_{-\infty}^\infty q_K(x,\sigma,\tc) f(x+\sigma\tc)e^{K(x+\sigma \tc) - K(x)} dx.
\label{Eq:defN2}
\end{align}
We begin with the analysis of $N_1(\sigma,\tc)$.
Since the support of $f_K(x)$ in question is $x<0$, we write $N_1(\sigma,\tc)$ as
\begin{align}
\frac{N_1(\sigma,\tc)}{2\sigma \nu^2\alpha(1-\alpha)} &=  \int_{-\infty}^{-2 u(\sigma)\tc} dx 
\left [ x(x+2\sigma \tc)\right ]^{\nu-1} e^{-\zeta_{11}(-x,\sigma,\tc)}\nonumber \\
& = \int_0^\infty dy \left [ y(y+2 \tc)\right ]^{\nu-1} e^{-\zeta_{11}(y + 2 u(\sigma)\tc,\sigma,\tc)},
\end{align}
where $u(\sigma) = \text{max}(0,\sigma)$,
we have changed variables to $y = -x- 2 u(\sigma) \tc$ to get the second line, and
\begin{align}
\zeta_{11}(z,\sigma,\tc) = z^\nu + (1-\omega_\sigma) \left [ (z-2\sigma\tc)^\nu - z^\nu \right ]
+ e^{- (z-\sigma\tc)^\nu}-e^{-z^\nu}.
\end{align}
If we again change variables to $z = y/\tc$, 
the above integral becomes
\begin{align}
\frac{N_1(\sigma,\tc)}{2\sigma \nu^2\alpha(1-\alpha)}
= \tc^{2\nu-1} \int_0^\infty \left [ z (z+2)\right ]^{\nu-1} e^{-\zeta_{12}(z,\sigma,\tc)} dy,
\end{align}
where 
$\zeta_{12}(z,\sigma,\tc)  = \zeta_{11}(\tc z +2 \tc u(\sigma),\sigma,\tc)$,
which is zero if $\tc = 0$.
Thus, the leading behavior of $N_1$ is
\begin{align}
\frac{N_1(\sigma,\tc)}{2\nu^2\alpha(1-\alpha)}
= \sigma \tc^{2\nu-1} \int_0^\infty \left [ y (y+2)\right ]^{\nu-1} dy + o(\tc^{2\nu-1}),
\label{Eq:N1}
\end{align}
where the integral is finite if $\nu < \frac{1}{2}$.
Thus, $N_1(1,\tc) + N_1(-1,\tc) = o(\tc^{2\nu-1})$ if $\nu$ is strictly smaller than
$\frac{1}{2}$.

When $\nu = \frac{1}{2}$, the integral in Eq.~\eqref{Eq:N1} is not
defined, which requires different approach for this case. To extract
the leading behavior, we performed integration by parts such that
\begin{align}
&\frac{N_1(\sigma,\tc)}{2\sigma \nu^2\alpha(1-\alpha)}
=  -2 \ln\left ( \sqrt{2\tc} \right ) \nonumber \\
&+ 2 \int_0^\infty dy \ln \left ( \sqrt{y} + \sqrt{y+2\tc} \right ) e^{-\zeta_{11}(y +2 u(\sigma)\tc,\sigma,\tc)} \frac{d\zeta_{11}}{dy} \nonumber \\
&=- \ln (2 \tc) + 2 \int_0^\infty \frac{e^{-\sqrt{x}}}{2\sqrt{x}} \ln \left (2 \sqrt{x} \right ) dx + o(1) \nonumber \\
&= -\ln \tc + \ln 2 - 2 \gamma + o(1),
\end{align}
where we have used (${\cal S}>0$)
\begin{align}
\frac{d}{dx} \ln \left ( \sqrt{x} + \sqrt{x+{\cal S}} \right )
= \frac{1}{2\sqrt{x(x+{\cal S})}},
\end{align}
and $\gamma \approx 0.5771$ is the Euler-Mascheroni constant.
Still, $N_1(1,\tc) + N_1(-1,\tc) = o(1)$ even for $\nu = \frac{1}{2}$.
Hence, $N_1$ does not contribute to the leading behavior of $\psi(\tc)$.

Now we move on to the analysis of $N_2$. We first write $N_2$ as
\begin{align}
\frac{N_2(\sigma,\tc)}{\sigma \omega_\sigma \nu^2}
&= \int_{-\infty}^{-u(\sigma)\tc} \left [ x(x+\sigma \tc) \right ]^{\nu-1}
e^{-\zeta_{21}(-x,\sigma,\tc)} dx \nonumber \\
&= \int_{0}^{\infty} \left [ y(y+ \tc) \right ]^{\nu-1}
e^{-\zeta_{21}(y+u(\sigma)\tc,\sigma,\tc)} dy,
\end{align}
where we have made a change of variables $y = -x - u(\sigma) \tc$ and
\begin{align}
\zeta_{21}(z,\sigma,\tc) =& (1-\omega_\sigma) \left [ z^\nu + \ln K(-z+2 \sigma \tc) \right ] \nonumber \\
&+ z^\nu + (z-\sigma \tc)^\nu + e^{-z^\nu} - e^{-(z+\sigma \tc)^\nu}
\end{align}
Note that $\ln K(z) = 0$ if $z>0$.
As above, the change of variables to $z = y/\tc$ gives
\begin{align}
\frac{N_2(\sigma,\tc)}{\sigma \omega_\sigma \nu^2}
= \tc^{2\nu-1} \int_0^\infty dz \left [ z (z+1) \right ]^{\nu-1} e^{-\zeta_{22}(z,\sigma,\tc)},
\end{align}
where
$\zeta_{22}(z,\sigma,\tc)  = \zeta_{21}(\tc z +\tc u(\sigma),\sigma,\tc)$ with 
the property $\zeta_{22}(z,\sigma,0) = 0$.
Thus the leading behavior of $N_2$ is
\begin{align}
N_2(\sigma,\tc) &\approx \sigma \omega_\sigma \nu^2 \tc^{2\nu-1} \int_0^\infty dz \left [ z (z+1) \right ]^{\nu-1} \nonumber \\
&= \sigma \omega_\sigma \nu^2 \tc^{2\nu-1} \frac{\Gamma(1-2\nu)\Gamma(\nu)}{\Gamma(1-\nu)},
\end{align}
which is valid for $\nu < \frac{1}{2}$. For $\nu = \frac{1}{2}$, integrating by parts
gives
\begin{align}
&\frac{N_2(\sigma,\tc)}{\sigma \omega_\sigma \nu^2}
= - \ln \tc \nonumber \\
&+ 2 \int_0^\infty dy \ln \left [\sqrt{y} + \sqrt{y+\tc} \right ] e^{-\zeta_{21}(y + u(\sigma)\tc,\sigma,\tc)} \frac{d\zeta_{21}}{dy} \nonumber \\
&= -\ln \tc +  2 \int_0^\infty dy \frac{\ln \left (2 \sqrt{y} \right ) }{\sqrt{y}} e^{-2\sqrt{y}} +o(1)\nonumber\\ 
&= -\ln \tc - 2 \gamma + o(1).
\end{align}
Since $\omega_1 - \omega_{-1} = 2 \alpha - 1$, we finally get
the leading behavior of $\psi(\tc)$ for $\nu < \frac{1}{2}$ as 
\begin{align}
\psi(\tc) &\approx (2\alpha-1)\nu^2 \tc^{2\nu-1} \frac{\Gamma(1-2\nu)\Gamma(\nu)}{\Gamma(1-\nu)},\nonumber\\
\int_0^\tc \psi(x) dx &= (2\alpha-1) \tc^{2\nu} \frac{\Gamma(1-2\nu)\Gamma(\nu+1)}{2 \Gamma(1-\nu)},
\end{align}
and for $\nu = \frac{1}{2}$ as
\begin{align}
\psi(\tc) &\approx \frac{2\alpha-1}{4} \left ( -\ln \tc  - 2 \gamma \right ),
\nonumber \\
\int_0^\tc \psi(x) dx &= -\frac{2\alpha-1}{4} \tc \ln (e^{2 \gamma-1} \tc).
\end{align}

\bibliographystyle{elsarticle-harv} 
\bibliography{Park}

\end{document}